\documentclass[12pt]{spieman}  % 12pt font required by SPIE;
\usepackage{amsmath,amsfonts,amssymb}
\usepackage{graphicx}
\usepackage{setspace}
\usepackage{tocloft}
\usepackage{float}
\usepackage{color}
\usepackage{soul}
\usepackage{multirow,multicol}
\usepackage{lineno}

\title{Simulations of expected signal and background of gamma-ray sources by large field-of-view detectors aboard CubeSats}
\author[a,b,*]{G\'abor Galg\'oczi}
\author[a,c,d,e,\#]{Jakub {\v R}{\'i}pa}
\author[f]{Riccardo Campana}
\author[e,g,c]{Norbert Werner}
\author[h]{Andr\'as P\'al}
\author[a,c,g]{Masanori Ohno}
\author[h]{L\'aszl\'o M\'esz\'aros}
\author[i]{Tsunefumi Mizuno}
\author[j]{Norbert Tarcai}
\author[g]{Kento Torigoe}
\author[g]{Nagomi Uchida}
\author[g]{Yasushi Fukazawa}
\author[g]{Hiromitsu Takahashi}
\author[k]{Kazuhiro Nakazawa}
\author[g]{Naoyoshi Hirade}
\author[g]{Kengo Hirose}
\author[k]{Syohei Hisadomi}
\author[l]{Teruaki Enoto}
\author[m]{Hirokazu Odaka}
\author[n]{Yuto Ichinohe}
\author[a]{Zsolt Frei}
\author[h]{L\'aszl\'o Kiss}
\affil[a]{E\"otv\"os University, Institute of Physics, P\'azm\'any P\'eter s\'et\'any 1/A, Budapest, Hungary, 1117}
\affil[b]{Wigner Research Centre, Konkoly-Thege Mikl\'os \'ut 29-33., Budapest, Hungary, 1121}
\affil[c]{MTA-E\"ot\"vos University Lend\"ulet Hot Universe Research Group, P\'azm\'any P\'eter s\'et\'any 1/A, Budapest, Hungary, 1117}
\affil[d]{Charles University, Faculty of Mathematics and Physics, Astronomical Institute, V Hole\v{s}ovi\v{c}k\'ach 2, Prague 8, Czech Republic, 180 00}
\affil[e]{Masaryk University, Faculty of Science, Department of Theoretical Physics and Astrophysics, Kotl\'a\v{r}sk\'a 2, Brno, Czech Republic, 611 37}
\affil[f]{ INAF - Astrophysical and Space Science Observatory (OAS), Via Gobetti 101, Bologna, Italy, 40129}
\affil[g]{Hiroshima University, School of Science, 1-3-1 Kagamiyama, Higashi-Hiroshima, Japan, 739-8526}
\affil[h]{Konkoly Observatory of the Hungarian Academy of Sciences, Konkoly-Thege ut 15-17, Budapest, Hungary, 1121}
\affil[i]{Hiroshima University, Hiroshima Astrophysical Science Center, 1-3-1 Kagamiyama, Higashi-Hiroshima, Japan, 739-8526}
\affil[j]{C3S Electronics Development LLC., K{\"o}nyves K\'alm\'an krt. 12-14., Budapest, Hungary, 1097}
\affil[k]{Nagoya University, Department of Physics, Furo-cho, Chikusa-ku, Nagoya, Aichi, Japan, 464-8602}
\affil[l]{Kyoto University, The Hakubi Center for Advanced Research and Department of Astronomy, Kyoto, Japan, 606-8302}
\affil[m]{University of Tokyo, Department of Physics, 7-3-1 Hongo, Bunkyo, Tokyo, Japan, 113-0033}
\affil[n]{Rikkyo University, Department of Physics, Nishi Ikebukuro 3-34-1, Toshimaku, Tokyo, Japan, 171-8501}

\cftpagenumbersoff{figure}
\cftpagenumbersoff{table} 
\begin{document}
\maketitle

\begin{abstract}
In recent years the number of CubeSats (U-class spacecrafts) launched into space has increased exponentially marking the dawn of the nanosatellite technology. In general these satellites have a much smaller mass budget compared to conventional scientific satellites which limits shielding of scientific instruments against direct and indirect radiation in space.

In this paper we present a simulation framework to quantify the signal in large field-of-view gamma-ray scintillation detectors of satellites induced by X-ray/gamma-ray transients, by taking into account the response of the detector. Furthermore, we quantify the signal induced by X-ray and particle background sources at a Low-Earth Orbit outside South Atlantic Anomaly and polar regions. Finally, we calculate the signal-to-noise ratio taking into account different energy threshold levels. Our simulation can be used to optimize material composition and predict detectability of various astrophysical sources by CubeSats.

We apply the developed simulation to a satellite belonging to a planned \textit{CAMELOT} CubeSat constellation. This project mainly aims to detect short and long gamma-ray bursts (GRBs) and as a secondary science objective, to detect soft gamma-ray repeaters (SGRs) and terrestrial gamma-ray flashes (TGFs). The simulation includes a detailed computer-aided design (CAD) model of the satellite to take into account the interaction of particles with the material of the satellite as accurately as possible.

Results of our simulations predict that CubeSats can complement the large space observatories in high-energy astrophysics for observations of GRBs, SGRs and TGFs. For the detectors planned to be on board of the \textit{CAMELOT} CubeSats the simulations show that detections with signal-to-noise ratio of at least 9 for median GRB and SGR fluxes are achievable.
\end{abstract}

% Include a list of up to six keywords after the abstract

% Include email contact information for corresponding author
{\noindent \footnotesize\textbf{*}G\'abor Galg\'oczi, \linkable{galgoczi@caesar.elte.hu} }
{\noindent \footnotesize\textbf{\#}Jakub {\v R}{\'i}pa, \linkable{jakub.ripa@ttk.elte.hu} }

\begin{spacing}{2}   % use double spacing for rest of manuscript

\section{Introduction}
\label{sect:intro}
Particle background is a considerable constraint for satellites, particularly those aiming to investigate the high-energy Universe. It is especially important for instruments without an anti-coincidence shield, e.g. for the increasingly large number of CubeSats which have recently been proposed for scientific missions.

A dedicated Geant4 \cite{geant4} software has been developed including the simulation of the optical light propagation inside the scintillators used as detectors. This way the detector response can be taken into account. In order to include the effects of scattering, photon conversion and other interactions happening between background particles, X-ray photons and the materials of the satellite, a computer-aided design (CAD) model of the whole satellite was included in the simulations.

The spectra of high energy photons and particles which contribute to the overall detected background were used as an input to the Geant4 simulations. These components of the external background include cosmic X-rays/$\gamma$-rays, cosmic ray particles, geomagnetically trapped particles and albedo (secondary) particles produced in the Earth's atmosphere.

In order to validate the background simulations, a set of dedicated experiments were carried out at the Hiroshima University in order to obtain the scintillator optical parameters (e.g. reflectivity of the surfaces and absorption length) which determine the position dependence of signal collection efficiency.

The developed simulation, spectra of the X/$\gamma$-rays and particle background as well as example spectra of high-energy photon transients were applied on one 3U CubeSat belonging to the planned Cubesats Applied for MEasuring and LOcalising Transients (\textit{CAMELOT}) constellation \cite{Werner2018, Pal2018, Ripa2018}. This simulation framework can be also helpful for other CubeSat and SmallSat missions with gamma-ray detectors in preparation by other teams, e.g. \textit{BurstCube} \cite{Smith2019, Perkins2017}, \textit{BlackCAT} \cite{Chattopadhyay2018}, Gravitational wave high-energy Electromagnetic Counterpart All-sky Monitor
(\textit{GECAM}) \cite{Zheng2019}, Gamma-Ray Integrated Detectors (GRID) \cite{Wen2019}, \textit{Glowbug} \cite{Grove2020}, High Energy Rapid Modular Ensemble of Satellites - Scientific Pathfinder (\textit{HERMES-SP}) \cite{Fuschino2019}, Satellite Polarimeter for High eNergy X-rays (\textit{SPHiNX}) \cite{Pearce2019}, \textit{SkyHopper}\footnote{\url{https://skyhopper.research.unimelb.edu.au}}, Space Industry Responsive Intelligent Thermal satellite, (\textit{SpIRIT})\footnote{\url{https://spirit.research.unimelb.edu.au}}.

The paper is organized as follows:
Sec.~\ref{sec:transients} describes the astrophysical sources whose detectability is investigated,
Sec.~\ref{sec:bck} overviews various background components,
Sec.~\ref{sec:camelot} describes the \textit{CAMELOT} satellites and the detector system,
Sec.~\ref{sec:validate} details the validation of Geant4 simulations and the calibration of the detector's optical parameters,
Sec.~\ref{sec:simulations} describes the performed Geant4 simulations,
Sec.~\ref{sec:sim_results} presents the results of Geant4 simulations and
Sec.~\ref{sec:conclusions} summarizes the conclusions.

\section{Expected X-ray/$\gamma$-ray Transient Sources}
\label{sec:transients}

The main scientific objective of the proposed \textit{CAMELOT} satellites is the detection of GRBs \cite{Ripa2019, Bagoly2019}. Short GRBs (sGRBs) originate from a merger of two neutron stars and possibly also from a merger of a neutron star and a black hole \cite{Vedrenne2009, Kouveliotou2012, Levan2018, Zhang2019}. The typical duration $T_{90}$ (the time during which the cumulative counts increase from 5\,\% to 95\,\%) of their prompt gamma-ray emission is $\lesssim 2$\,s in the observer frame \cite{Kouveliotou1993} and their gamma-ray energy flux peaks at $\sim 600$\,keV. Long GRBs (lGRBs) originate in the gravitational collapse of fast-spinning massive stars and their typical duration is $\gtrsim 2$\,s. The prompt spectra of lGRBs are on average softer then the sGRB spectra with their energy flux peaking at $\sim 200$\,keV.

\begin{figure*}[h]
	\centerline{
	\includegraphics[width=0.99\linewidth]{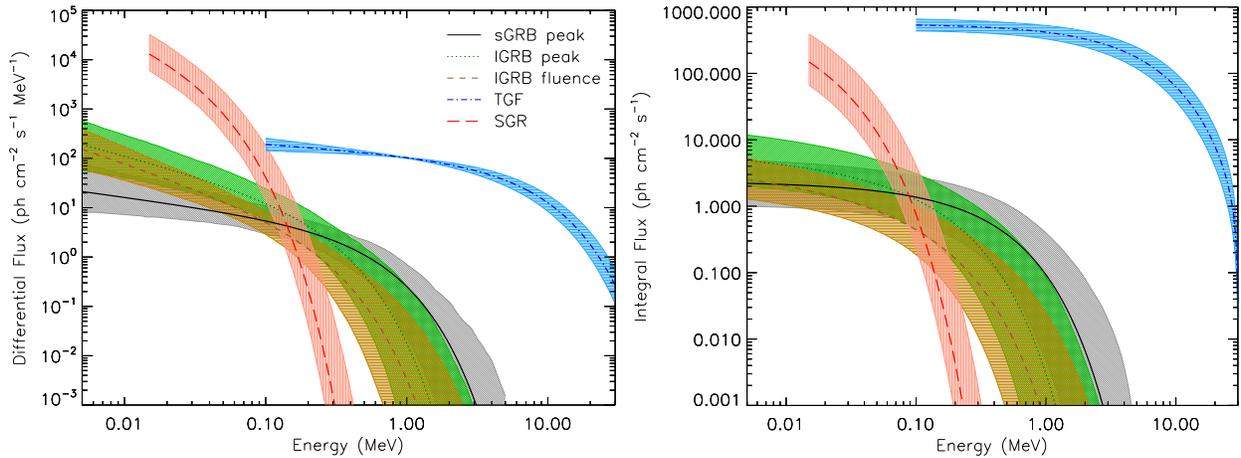}
	}
	\caption{Differential (left) and integral (right) spectra of typical GRBs, SGR and TGF. The black solid curve shows a typical peak spectrum of a sGRB. The black dotted curve shows a typical peak spectrum of a lGRB and the black dashed line shows a typical fluence spectrum of a lGRB accumulated over the duration of the burst. The blue solid curve shows an average spectrum of a TGF based on measurements from the \textit{AGILE} satellite. The red solid curve shows a typical spectrum of a burst from a SGR based on measurements from the Konus experiment. The shaded regions correspond to 68\,\% CL.
\label{fig:sources}}
\end{figure*}

The \textit{CAMELOT} satellites might be sensitive also to other astrophysical X-ray transients such as soft gamma repeaters with typical duration of individual peaks in their light curves $\sim0.2$\,s and with X-ray energy flux peaking at around 20\,keV (see Sec.~\ref{sec:SGR}). Note that a hard component above 100\,keV has been observed as well \cite{Mereghetti2008, Yamaoka2017}. Also the gamma-ray phenomena produced in the Earth's atmosphere during thunderstorms \cite{Lindanger2020} called terrestrial gamma-ray flashes might be observed. These events are typically shorter than 1\,ms and have gamma-ray spectra reaching energies of several MeV \cite{Dwyer2012}. The following subsections describe in detail the fluxes expected from these sources. Fig.~\ref{fig:sources} summarizes the spectra of the X-ray/$\gamma$-ray transient sources which we study in this paper.

\subsection{Typical Short and Long Gamma-Ray Burst Spectra}
\label{sec:GRB}

Since the main objective of the \textit{CAMELOT} mission is the detection of GRBs, we run Monte Carlo (MC) simulations using the typical spectra of sGRBs and lGRBs in order to estimate the expected signal-to-noise ratio. The typical spectra were constructed using the \textit{Fermi} GBM Burst Catalog (FERMIGBRST\footnote{\url{https://heasarc.gsfc.nasa.gov/W3Browse/fermi/fermigbrst.html}}). For detailed information about the catalog see Ref.~\citenum{Gruber2014, vonKienlin2014, NarayanaBhat2016, vonKienlin2020}. For sGRBs, we used the so called peak flux spectrum which is accumulated over the peak of the GRB (typically over 64\,ms or 1024\,ms). In case of lGRBs, we used the peak flux spectrum and the so called fluence spectrum, which is accumulated over the whole duration of the burst.

Note that the 64, 256 and 1024\,ms timescales of measured peak fluxes reported in the catalog do not match all the trigger timescales used by \textit{Fermi}/GBM. The triggering system employs 120 possible sets of trigger algorithms (not all actively employed at a time and approximately 60 trigger algorithms are currently active\cite{vonKienlin2020}) consisting of eight set of energy bands, ten time scales 16, 32, 64, 128, 256, 512, 1024, 2048, 4096, 8192\,ms and different time offsets for two phases of selected time interval. Most frequently, GBM triggers on 5 time scales from 16 to 4096\,ms \cite{NarayanaBhat2016}.

The typical spectra of sGRB and lGRB were constructed in the following way. First, we checked what was the most common best fit spectral model in the catalog. For peak flux spectra of sGRBs it was the power law model (PL). For the peak flux and fluence spectra of lGRBs it was the Comptonized model (CPL, exponential cutoff power law). For detailed information about the different spectral models see Ref.~\citenum{Gruber2014}.

Although PL model was the most frequent one for peak spectra of sGRBs in the FERMIGBRST catalog, we use the second most frequent model, i.e. CPL. The reason is that sGRBs dim in peak flux are most frequently best fit by PL whereas brighter sGRBs are most frequently fit by CPL model. A likely explanation is that short GRB have Comptonized spectra, and that weak sGRBs produce insufficient signal in the instrument to distinguish the models.

Then in case of the peak spectra, we used the median best fit spectral parameters and then we tuned the normalizations $A$ of the spectra to obtain the values of the integral fluxes in the range of $10-1000$\,keV equal to the median 1024\,ms, 256\,ms and 64\,ms timescale peak fluxes obtained from the catalog. The median peak fluxes for sGRBs for 1024\,ms, 256\,ms and\,64 ms time scales and in the $10-1000$\,keV range are 2.0, 4.8 and 7.5 ph\,cm$^{-2}$s$^{-1}$, respectively. Note that the median power law index for PL model for sGRBs peak spectra is $\alpha=-1.4$ which is unphysical \cite{Goldstein2016} and that is also the reason why we use CPL model for sGRB spectra. In case of lGRBs the median peak fluxes for 1024\,ms, 256\,ms and\,64 ms time scales and in the same energy range are 4.1, 5.2 and 6.7 ph\,cm$^{-2}$s$^{-1}$, respectively. For the fluence spectra of lGRBs, we used the median best fit spectral parameters, including the normalization, from the catalog. The obtained spectral parameters are in Table~\ref{tab:grb_param}. The pivot energy $E_\textrm{piv}$ is fixed at 100\,keV. Figure~\ref{fig:sources} shows the typical GRB spectra with shaded regions corresponding to 68\,\% CL. These 68\,\% CL were obtained from the measured spectral parameters separately for short and long GRBs in the FERMIGBRST catalog and using the CPL spectral model.

\begin{table}[h]
\caption{Spectral parameters of typical GRB spectra.}
\begin{center}
\begin{tabular}{|c|c|c|c|c|c|c|c|}
\hline
\rule[-1ex]{0pt}{3.5ex} GRB  & Spec.& $A_{1024}$ & $A_{256}$ & $A_{64}$ & $A$    & $\alpha$ & $E_\mathrm{peak}$ \\
\rule[-1ex]{0pt}{3.5ex} type & type &            &           &          &        &          &             (keV) \\
\hline
\rule[-1ex]{0pt}{3.5ex} sGRB & pflx & $0.0068^{+0.0090}_{-0.0035}$ & $0.016^{+0.021}_{-0.008}$ & $0.025^{+0.033}_{-0.013}$ & ---                            & $-0.38^{+0.34}_{-0.30}$ & $669^{+574}_{-350}$ \\ 
\rule[-1ex]{0pt}{3.5ex} lGRB & pflx & $0.020^{+0.036}_{-0.012}$    & $0.026^{+0.045}_{-0.015}$  & $0.033^{+0.058}_{-0.019}$ & ---                            & $-0.75^{+0.37}_{-0.30}$ & $235^{+269}_{-117}$ \\ 
\rule[-1ex]{0pt}{3.5ex} lGRB & flnc & ---                               & ---                             & ---                            & $0.009^{+0.010}_{-0.004}$ & $-0.96^{+0.37}_{-0.31}$ & $183^{+250}_{-89}$  \\ 
\hline
\end{tabular}
\end{center}
The spectral parameters are for the peak flux spectra (pflx) and the fluence spectra (flnc) of typical short and long GRBs. The normalizations $A_{1024}$, $A_{256}$ and $A_{64}$ are for the 1024\,ms, 256\,ms and 64\,ms timescale peaks, respectively, $A$ is the normalization for the fluence spectrum and all normalizations are in units of ph\,cm$^{-2}$s$^{-1}$keV$^{-1}$. Parameters $\alpha$ and $E_\mathrm{peak}$ are respectively the power law index and the peak energy of the Comptonized model. The uncertainties correspond to 68\,\% CL.
\label{tab:grb_param}
\end{table}

\subsection{Soft Gamma Repeaters}\label{sec:SGR}

Soft gamma repeaters and anomalous X-ray pulsars (AXPs) are believed to be neutron stars with extremely strong magnetic fields of up to $B\sim10^{14}-10^{15}$\,G called ``magnetars" \cite{Duncan1992, Thompson1995, Thompson1996}. For reviews see Refs.~\citenum{Lewin2006, Kaspi2007, Mereghetti2008, Enoto2019}. First observations date to 1979 \cite{Mazets1979} with several magnetar bursts detected up to now, see e.g. Refs.~\citenum{Kouveliotou1998, Mereghetti2005a, Molkov2005, Yamaoka2017}. For example a giant flare of magnetar SGR 1806-20 on 27 December, 2004 was observed by several satellites \cite{Hurley2005, Mazets2005, Mereghetti2005b, Palmer2005, Terasawa2005} and it was so bright that it saturated detectors.

SGR giant flares are rare and it is essential for their better understanding to observe and monitor all of them. The all sky coverage provided by future networks of nano-satellites will ensure that all future SGR outbursts will be detected and their behaviour will be monitored.

The SGR spectra are in soft gamma-ray region and well represented by a single blackbody (BB) or two-temperature BB model \cite{Yamaoka2017}. Above $\sim30$\,keV the spectra are well modeled by optically thin thermal bremsstrahlung (OTTB) \cite{Lewin2006}.

In our simulations we analyze the response to regular SGR bursts. For an example spectrum of a regular SGR we utilize the Konus catalog of SGRs detected from 1978 to 2000 \cite{Aptekar2001}. The catalog contains bursts from SGR 0526-66, 1627-41, 1801-23, 1806-20 and 1900+14 observed using detectors on board Venera 11--14, Wind, and Kosmos 2326 spacecrafts. The SGR photon spectra in the catalog are modeled by optically thin thermal bremsstrahlung using:

\begin{equation}
F(E)=AE^{-1}e^{-\frac{E}{kT}},
\end{equation}
where $kT$ (keV) is the spectral parameter, $E$ (keV) is the photon energy and $A$ (ph\,cm$^{-2}$s$^{-1}$) is the normalization factor.

\begin{figure*}[h]
	\centerline{
	\includegraphics[width=0.99\linewidth]{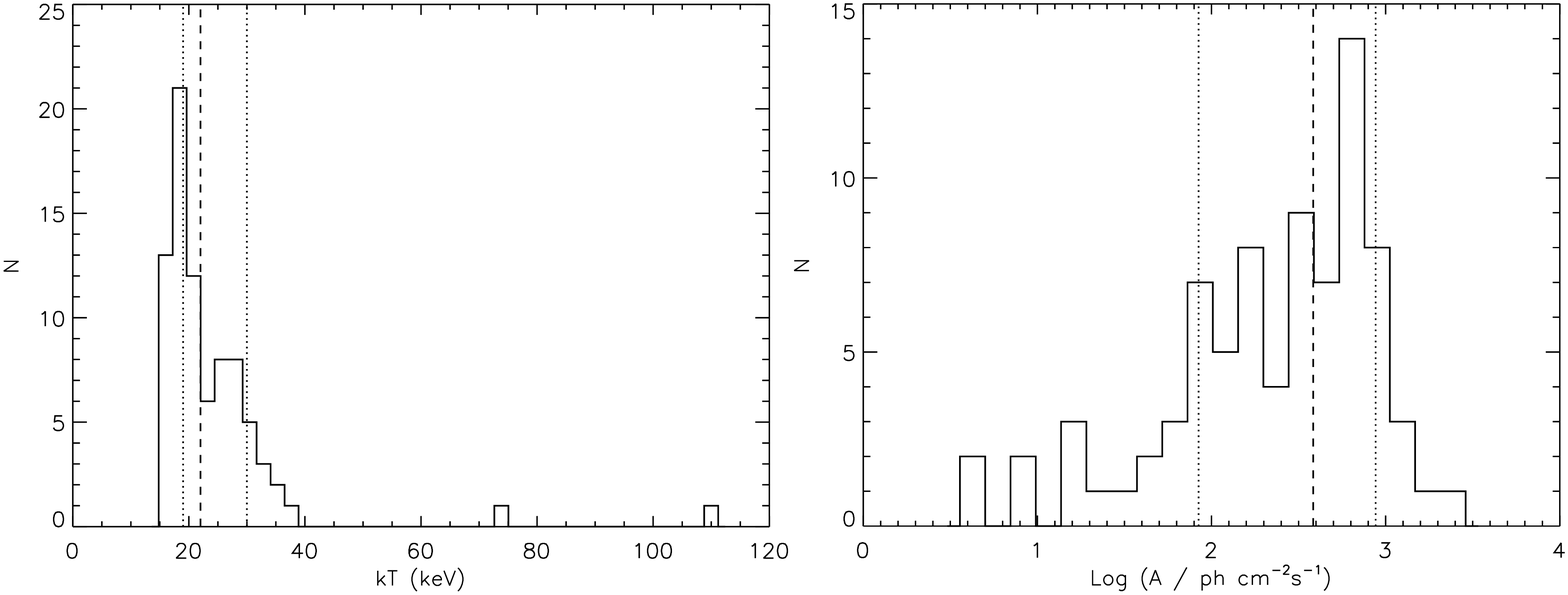}
	}
	\caption{Distribution of the spectral parameter $kT$ (left) and normalization $A$ (right) from the Konus SGR catalog \cite{Aptekar2001} for 81 spectral fits. The dashed lines mark the median values and the regions delimited by dotted lines correspond to 68\,\% CL.
\label{fig:sgr_parameters}}
\end{figure*}

We took 81 spectral fits from the catalog and derived the normalization factor $A$ from each spectrum. Figure~\ref{fig:sgr_parameters} shows the distribution of the $kT$ parameter and the normalization $A$. The median values are $kT=22^{+8}_{-3}$\,keV and $A=384^{+493}_{-300}$\,ph\,cm$^{-2}$s$^{-1}$. The uncertainties correspond to 68\,\% CL. The typical duration of SGR bursts in the catalog is $\sim0.2$\,s. We use these parameters for an example SGR spectrum used in our MC simulations in the energy range of $15-500$\,keV. The spectrum is shown in Figure~\ref{fig:sources} with the shaded region corresponding to 68\,\% CL which was calculated using MC simulations and the aforementioned values of $kT$ and $A$ parameters with their uncertainties.

\subsection{Terrestrial Gamma-Ray Flashes}\label{sec:TGF}

Terrestrial gamma-ray flashes, brief bright bursts of multi-MeV gamma-rays, which are believed to be emitted by thunderclouds and generated, via bremsstrahlung, by the relativistic runaway electrons accelerated by electric fields in the atmosphere \cite{Dwyer2012}. They were discovered by the Burst and Transient Source Experiment (BATSE) aboard Compton Gamma-ray Observatory (\textit{CGRO}) \cite{Fishman1994}. Plentiful observations have also been provided by other astrophysical instruments such as the Reuven Ramaty High Energy Solar Spectroscopic Imager (\textit{RHESSI}) \cite{Smith2005, Grefenstette2009}, the Gamma Ray Burst Monitor (GBM) on board the \textit{\textit{Fermi}} satellite \cite{Briggs2010, Fishman2011, Roberts2018}, and the Astrorivelatore Gamma ad Immagini Leggero (\textit{AGILE}) satellite \cite{Marisaldi2011, Marisaldi2014, Marisaldi2015}. The \textit{AGILE} satellite was launched to 550\,km altitude with inclination of $2.5^\circ$. The Mini-Calorimeter (MCAL) is composed of 30 CsI(Tl) scintillator bars with each crystal of dimension of $15\times23\times375$\,mm$^3$ giving the detector a sensitivity from 300\,keV to 200\,MeV and the effective area of $200-1200$\,cm$^2$ \cite{Labanti2009}. Although this makes \textit{AGILE}/MCAL a more sensitive instrument for TGF detection than what is foreseen for \textit{CAMELOT}, which is expected to be launched to polar LEO with detectors composed of 5\,mm thick CsI(Tl) crystals, we use the \textit{AGILE}/MCAL observations of TGFs as a reference for our simulations because the mission accumulated a large and high quality TGF database in orbit. For their TGF observations see the 3rd \textit{AGILE} TGF catalog \cite{Lindanger2020, Maiorana2020} and the corresponding online catalog\footnote{\url{https://www.ssdc.asi.it/mcal3tgfcat/}}.

The duration of TGFs is typically below 1\,ms with the peak of the distribution around $100-200\,\mu$s \cite{Marisaldi2011, Marisaldi2014, Marisaldi2015}. \textit{AGILE} measurements show that the cumulative spectrum of 228 single-pulse TGFs in the range $0.4-30$\,MeV can be fitted by a power law with exponential cutoff \cite{Marisaldi2014}:

\begin{equation}
F(E)=K\left ( \frac{E}{1\,\text{MeV}} \right )^{-\alpha}e^{-\frac{E}{E_\mathrm{C}}},
\end{equation}

where $\alpha=0.20^{+0.12}_{-0.13}$ and $E_\mathrm{C}=5.5^{+0.7}_{-0.6}$\,MeV. The cumulative spectrum is a rough approximation because of the effects due to atmospheric absorption from different source regions and due to the direction-dependent detector response which in the cumulative spectrum are smeared out. For analysis of instrumental effects and their impact on energy spectra see \cite{Marisaldi2019}.

The fluence distribution can be represented with a power law with an index of $-$2.2 to $-$2.4 \cite{Marisaldi2014, Tierney2013}. Therefore, there is no typical observed TGF fluence. However, for a reference we can consider a typical TGF fluence at the threshold level of \textit{AGILE} which is around 0.05\,ph\,cm$^{-2}$ over the full energy range of \textit{AGILE}/MCAL $0.3-30$\,MeV and this typical fluence is emitted over an average duration $T_{50}$ of less than $100\,\mu s$ (M. Marisaldi, private communication) \cite{Marisaldi2015}. The normalization $K$ corresponding to this fluence integrated over $0.3-30$\,MeV and within $100\,\mu s$ is $K=123\,$ph\,cm$^{-2}$s$^{-1}$MeV$^{-1}$. We assume this normalization in our MC simulations. The spectrum extrapolated to 100\,keV is shown in Figure~\ref{fig:sources} with the shaded region corresponding to 68\,\% CL which was calculated using MC simulations and the aforementioned values of $\alpha$ and $E_\mathrm{C}$ parameters with their uncertainties. We assume that the published uncertainties of $\alpha$ and $E_\mathrm{C}$ correspond to 68\,\% CL. Normalization $K$ was kept fixed because it corresponds to the fluence detection threshold of TGFs by \textit{AGILE}.

\section{Components of Cosmic, Albedo and Trapped Particle Radiation}
\label{sec:bck}

\begin{figure*}[h]
	\centerline{
	\includegraphics[width=0.99\linewidth]{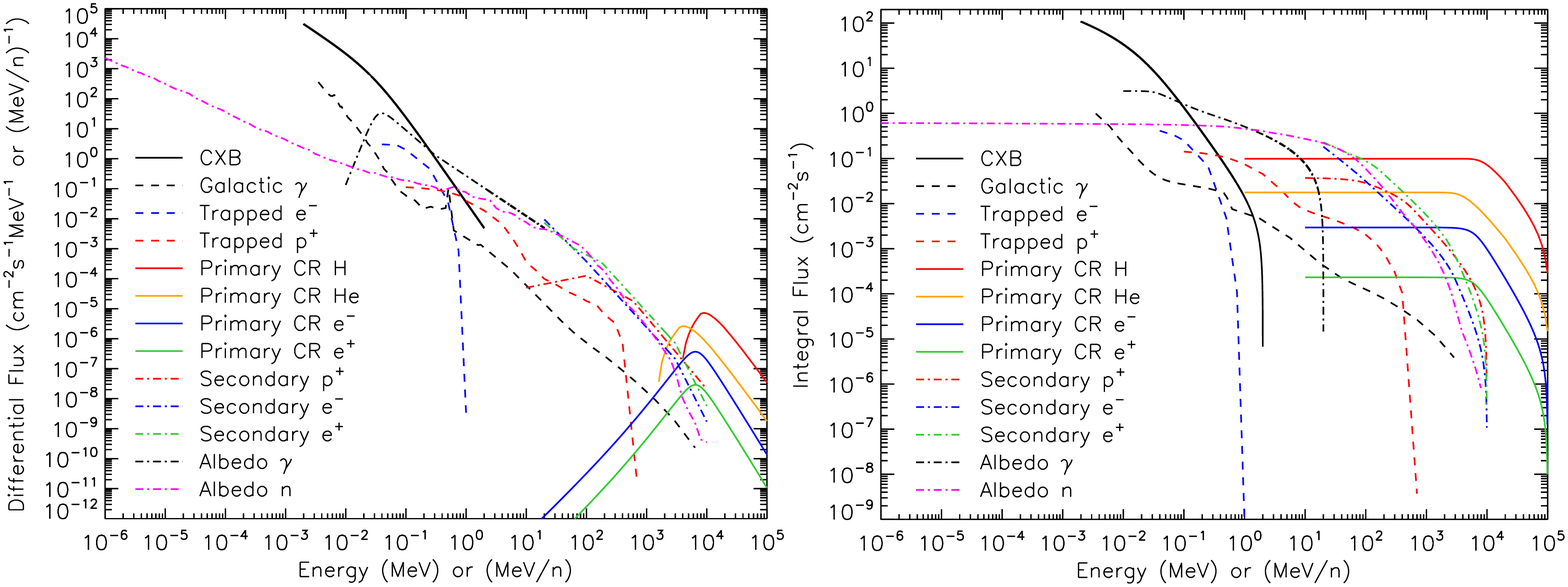}
	}
	\caption{An overview of various background components. \textit{Left:} The differential photon flux multiplied by the solid angles of the incident radiation valid for the expected altitude of the satellite of 500\,km and for all-sky field-of-view. Trapped particle, primary CR H and He fluxes are for the orbital inclination $i=20^\circ$. Primary CR e$^-$ and e$^+$ fluxes are for the geomagnetic latitude $\theta_\mathrm{M}=29.6^\circ$. Secondary p$^+$ flux was obtained from the combination of data for $0.3 \textrm{ rad} \leq \theta_\textrm{M} \leq 0.4 \textrm{ rad}$ and for $1.0 \leq \textrm{L-shell} \leq 1.7$, where L-shell is the McIlwain L-parameter. Secondary e$^-$ and e$^+$ fluxes were obtained from the combination of data for $0 \textrm{ rad} \leq \theta_\textrm{M} \leq 0.3 \textrm{ rad}$ and for $1.0 \leq \textrm{L-shell} \leq 1.2$. Albedo $\gamma$ flux is for $i=20.6^\circ$. Albedo n is for cutoff rigidity $R_\mathrm{cut}=5$\,GV or $\theta_\mathrm{M}=37^\circ$. Details are described in Sec.~\ref{sec:bck}. \textit{Right:} The integral photon flux for the same models also multiplied by the same solid angles. \label{fig:spectra_all}}
\end{figure*}

There are several external background components at Low-Earth Orbits (LEO) which need to be considered in a study of the expected detected background count rate by an instrument with large field-of-view (FOV). The various components include extragalactic gamma-rays, cosmic-ray particles, secondary gamma-rays and particles produced in the Earth's atmosphere and the Galactic gamma emission. For overview see publications Refs.~\citenum{Jursa1985, Armstrong1992, Gehrels1992, Dean2003, Barth2003, Mizuno2004, Zombeck2007, Fioretti2012, Campana2013, Xapsos2013, Zabori2018, Cumani2019, Mate2019} and technical notes Refs.~\citenum{ECSS-E-ST-10-04C, LAT-TD-08316-01}. The following subsections describe each component in detail and Fig.~\ref{fig:spectra_all} shows a comparison of the in-orbit expected background fluxes which we use in our Geant4 simulations involving the mass model of one 3U \textit{CAMELOT} CubeSat and its detectors.

\subsection{Cosmic X-ray/$\gamma$-ray Background}\label{sec:cxb}
The cosmic X-ray/$\gamma$-ray background (CXB) was discovered by a sounding rocket in 1962 \cite{Giacconi1962}.
It is nearly isotropic emission detected over a wide range of energies from few keV to few 100\,GeV \cite{Kasturirangan1972, Schwartz1974, Schoenfelder1977, Fichtel1978b, Schoenfelder1980, Boldt1981, Gehrels1992, Horstman1975, Sreekumar1998, Dean2003, Mizuno2004, Churazov2007, Frontera2007, Zombeck2007, Ajello2008, Ajello2009, Moretti2009, Turler2010, Campana2013, Ackermann2015}.
It is composed of high-energy emission from various extragalactic sources (active galactic nuclei, quasi-stellar objects, supernovae Ia, galaxy clusters, starburst galaxies, X-ray binaries, hot intergalactic gas) \cite{Bagoly1988, Meszaros1988, Bi1991, Jahoda1991, Shanks1991, Fabian1992, Comastri1995, Zdziarski1996, Sreekumar1998, Dean2003, Brandt2005, Ajello2008, Cappelluti2012, Helgason2014, Cappelluti2017, Ma2018}.
Some authors also argue that the diffuse $\gamma$-ray radiation originates in Cosmic Microwave Background \cite{Penzias1965} being inverse Compton scattered on cosmic-ray electrons \cite{Moskalenko2000, Dar2001}.

There are several empirical models used to describe the measured flux \cite{Kasturirangan1972, Gehrels1992, Gruber1999, Ajello2008, Cumani2019}. In our simulations we compare two models: the model introduced by Gruber et al. (1999) \cite{Gruber1999} and the one derived by Ajello et al. (2008) \cite{Ajello2008} (see Sec.~\ref{sec:sim_results}).

The Gruber et al. (1999) model \cite{Gruber1999} fits the low-energy as well as the high-energy part of the CXB measurements obtained by \textit{HEAO-1} \cite{Rothschild1979}, \textit{CGRO}/COMPTEL and \textit{CGRO}/EGRET \cite{Gehrels1993} instruments across a wide energy range spanning from 3\,keV to 100\,GeV. This empirical model is used as a standard in modeling of the CXB flux for planning space missions.
The differential photon flux $F(E) \equiv dN/dE$ in units of ph\,cm$^{-2}$s$^{-1}$sr$^{-1}$keV$^{-1}$ is:\\
for energies $E=3-60$\,keV
\begin{equation}
F(E)=7.877\left ( \frac{E}{1\,\mathrm{keV}} \right )^{-1.29}e^{-\frac{E}{41.13\,\mathrm{keV}}}
\end{equation}
and for energies $E>60$\,keV it is
\begin{equation}
F(E)=
\frac{0.0259}{60}\left ( \frac{E}{60\,\mathrm{keV}} \right )^{-6.5} +
\frac{0.504}{60}\left ( \frac{E}{60\,\mathrm{keV}} \right )^{-2.58} +
\frac{0.0288}{60}\left ( \frac{E}{60\,\mathrm{keV}} \right )^{-2.05}.
\end{equation}

Ajello et al. (2008)\cite{Ajello2008} derived a CXB model which is in a good agreement with measurements from \textit{Swift}/BAT\cite{Ajello2008}, \textit{HEAO-1}\cite{Gruber1999}, \textit{INTEGRAL}\cite{Churazov2007}, \textit{BeppoSAX}\cite{Frontera2007} instruments and other missions in the 2\,keV -- 2\,MeV energy range. The differential photon flux $F(E) \equiv dN/dE$ in units of ph\,cm$^{-2}$s$^{-1}$sr$^{-1}$keV$^{-1}$ is:
\begin{equation}\label{eq:Ajello2008}
F(E) = \frac{C}{(E/E_\mathrm{B})^{\Gamma_1} + (E/E_\mathrm{B})^{\Gamma_2}},
\end{equation}

where the parameters with 1\,$\sigma$ errors are $C= (10.15\pm0.80)\times 10^{-2}$, $\Gamma_1=1.32\pm0.018$, $\Gamma_2=2.88\pm0.015$ and $E_\mathrm{B}=29.99\pm1.1$\,keV.

\begin{figure*}[h]
	\centerline{
	\includegraphics[width=0.99\linewidth]{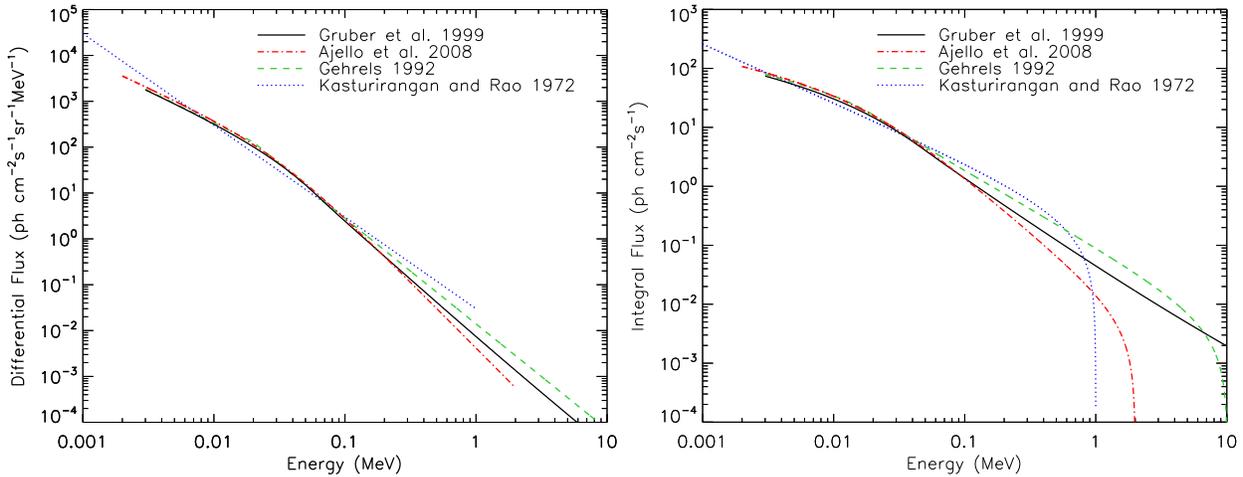}
	}
	\caption{The CXB spectra for different models. \textit{Left:} The differential photon flux. \textit{Right:} The integral photon flux for the same models. The integral flux is multiplied by a solid angle of the radiation illuminating the satellite at 500\,km altitude. \label{fig:cxb_spectra}}
\end{figure*}

The CXB flux is omnidirectional and for 500\,km altitude it irradiates a satellite from a solid angle of 8.64\,sr (3.93\,sr are occulted by the Earth). The CXB spectra for different models, including the two previously discussed, are shown in Figure~\ref{fig:cxb_spectra}. The integral flux, i.e. the integrated flux for energies above a given energy threshold, for the Gruber et al. (1999) model \cite{Gruber1999} and for $E>10$\,keV is 30.3\,ph\,cm$^{-2}$s$^{-1}$ whereas for the Ajello et al. (2008) model the integral flux at the same low-energy threshold is 33.7\,ph\,cm$^{-2}$s$^{-1}$.

Concerning the \textit{CAMELOT} CubeSats, the detectors are 5\,mm thick CsI(Tl) scintillators (as described in Sec.~\ref{sec:camelot}) with effective area having maximum at $\sim100$\,keV (see Figure~\ref{fig:eff_area}). The scintillator is relatively transparent to gamma-rays above 2\,MeV. For example the effective area at 1\,MeV is a factor of about 7 lower than at 100\,keV. Although some gamma-rays can cause pair-production or Compton scatter in the material of the satellite and then lower-energy gamma-rays can reach the scintillator, the CXB flux above 2\,MeV is much lower than at few tens of keV or at 100\,keV. Therefore, the high-energy gamma-ray component (above few MeV) included in the Gruber et al. (1999) model \cite{Gruber1999} is not essential for the \textit{CAMELOT}'s detectors. 

Ajello et al. (2008)\cite{Ajello2008} discusses that the normalization of the \textit{Swift}/BAT CXB spectrum at 30\,keV (CXB peak) is $\sim 8$\,\% higher than the \textit{HEAO-1}\cite{Gruber1999} measurement and consistent with the \textit{INTEGRAL}\cite{Churazov2007} one. Also, the \textit{HEAO-1} measurement has 10\,\% precision at the CXB peak. Therefore, we use both CXB models to simulate the expected detected background by \textit{CAMELOT}, however, for a more detailed analysis, e.g. detected count rate as a function of energy threshold, we choose a more conservative approach and use the Ajello et al. (2008) model which gives $\sim 9$\,\% higher integral flux in the energy range of $20-500$\,keV, which is approximately the sensitivity range of \textit{CAMELOT}'s detectors.

\subsection{Galactic Emission}\label{sec:galactic_gamma}
The Galactic gamma emission \cite{Melia2009, Schonfelder2001} which consists of diffuse continuum \cite{Strong2000} and resolved sources has been widely observed by many instruments, e.g. by \textit{SAS-2} \cite{Fichtel1975, Fichtel1978a, Fichtel1978b, Bertsch1993}, \textit{OSO-3} \cite{Kraushaar1972}, \textit{COS B} \cite{Mayer-Hasselwander1982, Bertsch1993}, \textit{INTEGRAL} \cite{Krivonos2007, Churazov2007, Porter2008, Bouchet2008, Turler2010, Krivonos2010, Krivonos2015} satellites, COMPTEL, EGRET and OSSE instruments \cite{Hunter1997, Sreekumar1998, Strong1999, Strong2004, Prantzos2011} aboard the \textit{CGRO} satellite, \textit{Fermi}/LAT \cite{Ackermann2015, Strong2011, Abdo2010, Ajello2016}, \textit{RXTE}/PCA \cite{Revnivtsev2006} and \textit{Swift}/BAT \cite{Oh2018} instruments.

For our MC simulations (Sec.~\ref{sec:sim_results}) we took the X-ray/gamma-ray fluxes of the inner Galactic region from Figures 10 and 11 as published in Ref.~\citenum{Turler2010} in $EF_\mathrm{E}$ flux density representation. Ref.~\citenum{Turler2010} summarizes measurements from \textit{RXTE}/PCA\cite{Krivonos2007}, \textit{INTEGRAL}/SPI\cite{Bouchet2008}, \textit{INTEGRAL}/IBIS\cite{Krivonos2007, Turler2010},
\textit{CGRO}/COMPTEL\cite{Porter2008} and \textit{CGRO}/EGRET\cite{Porter2008} instruments and show the fluxes renormalized to the central radian of the Milky Way defined by $|l|<30^\circ$ and $|b|<15^\circ$.

\begin{figure*}[h]
	\centerline{
	\includegraphics[width=0.99\linewidth]{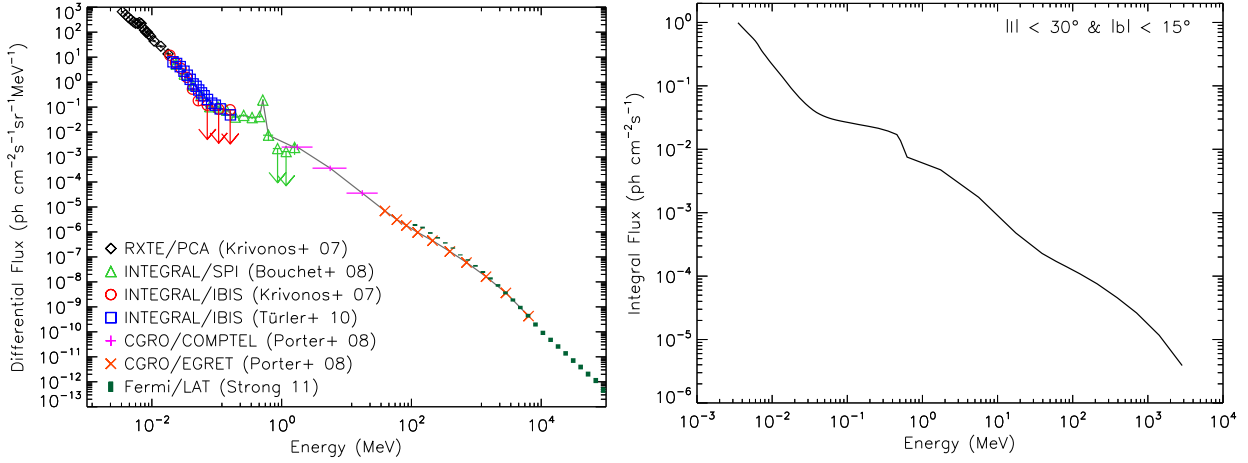}
	}
	\caption{The spectra of the inner Galaxy emission. \textit{Left:} The differential photon flux observed by \textit{RXTE}/PCA\cite{Krivonos2007}, \textit{INTEGRAL}/SPI\cite{Bouchet2008}, \textit{INTEGRAL}/IBIS\cite{Krivonos2007, Turler2010}, \textit{CGRO}/COMPTEL\cite{Porter2008} and \textit{CGRO}/EGRET\cite{Porter2008} for the region defined by $|l|<30^\circ$, $|b|<15^\circ$ and \textit{Fermi}/LAT\cite{Strong2011} for the region defined by $|l|<30^\circ$, $|b|<10^\circ$. The gray solid line marks the flux taken for our MC simulations. \textit{Right:} The integral photon flux. The integral flux is multiplied by 0.542\,sr solid angle of the inner Galaxy region. \label{fig:galactic_spectra}}
\end{figure*}

The emission from the inner Galactic region irradiates a satellite from the solid angle of 0.542\,sr. The differential and integral photon spectra are shown in Figure~\ref{fig:galactic_spectra}. The integral flux for energy $E>10$\,keV is 0.2\,ph\,cm$^{-2}$s$^{-1}$. However, it should be noted that the Galactic emission is not spatially uniform and has a brightness structure peaked at the Galactic center (see e.g. Ref.~\citenum{Mayer-Hasselwander1982}).
Figure~\ref{fig:galactic_spectra} also shows measurements done by \textit{Fermi}/LAT taken from Figure~4 of Ref.~\citenum{Strong2011} for smaller region of $|l|<30^\circ$ and $|b|<10^\circ$. We do not include these \textit{Fermi}/LAT measurements in our MC simulations because the region of the inner Galaxy is not exactly the same as the one used for the other aforementioned data sets. This does not effect our results because the photon flux at these very high energies is very small.

\subsection{Trapped Particles}\label{sec:trapped}
The fluxes of the geomagnetically trapped electrons and protons inside the inner van Allen radiation belt \cite{vanAllen1958} contribute to the overall detected instrumental background and they are especially important when a satellite at LEO passes the polar regions \cite{Svertilov2018, Panasyuk2018} or the South Atlantic Anomaly (SAA). Details about the Earth's radiation environment can be found in, for example, Ref.~\citenum{Roederer1970, Jursa1985, Kivelson1995, Walt2005, Zombeck2007, Qin2014, Adriani2015, Benghin2018, Shprits2018}.

Several models describing the fluxes of the trapped particles around the Earth based on measurements from tens of space missions have been developed over last decades, e.g. the National Aeronautics and Space Administration's (NASA) AE8 \cite{Vette1991a, Vette1991b} and AP8 \cite{Sawyer1976, Vette1991b} models, European Space Agency's (ESA) AE-8 update ESA-SEE1 \cite{Vampola1996} model, or model based on the measurements from Proton/Electron Telescope (PET) onboard the Solar, Anomalous, and Magnetospheric Particle Explorer (\textit{SAMPEX}) satellite - the \textit{SAMPEX}/PET PSB97 model \cite{Heynderickx1999}.

In our on-board background simulations (Sec.~\ref{sec:sim_results}) we employ the fluxes of trapped electrons and protons prescribed by the recent AE9 and AP9 models \cite{Ginet2013, Johnston2014, Johnston2015, OBrien2018} as they are implemented in ESA's SPace ENVironment Information System (SPENVIS\footnote{\url{www.spenvis.oma.be}}). SPENVIS is an Internet interface to models of the space environment and its effects, developed by a consortium led by the Royal Belgian Institute for Space Aeronomy (BIRA-IASB). The AE9/AP9 models are based on 33 satellite data sets from 1976 to 2011 and they are provided by the U.S. Air Force Research Laboratory (AFRL) in their software package \footnote{\url{https://www.vdl.afrl.af.mil/programs/ae9ap9}}.

The AE8/AP8 models available in SPENVIS or in the AFRL package do not compute fluxes lower than 1\,particle\,cm$^{-2}$s$^{-1}$, whereas the AE9/AP9 models provide fluxes below 1\,particle\,cm$^{-2}$s$^{-1}$. That is important for our purpose, because we want to estimate the detected background count rate in the regions outside SAA and polar regions. However, the current version of the AP9/AE9 model provided in SPENVIS is recommended for evaluation purposes only and there have been reported discrepancies between the AE8/AP8 and the AE9/AP9 models, e.g. see Ref.~\citenum{Pich2017} and references therein.

\begin{figure*}[h]
	\centerline{
	\includegraphics[width=0.99\linewidth]{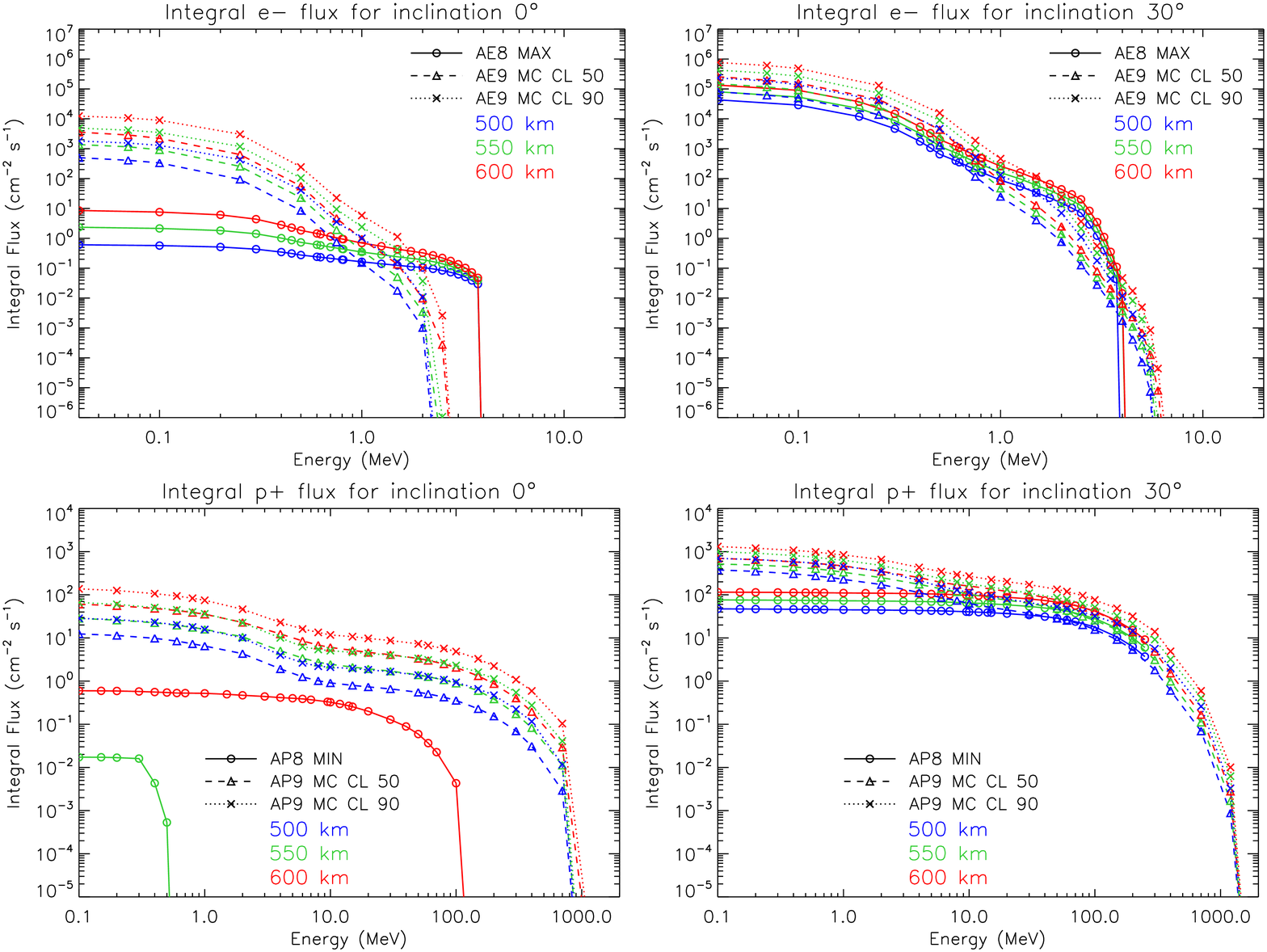}
    }
	\caption{Comparison of orbit-averaged integral fluxes of trapped electron models AE8 MAX (solar maximum), AE9 50\,\% and 90\,\% confidence levels (CL); and trapped proton models AP8 MIN (solar minimum), AP9 50\,\% and 90\,\% CL for different altitudes and inclinations.\label{fig:trapped_models_comparison}}
\end{figure*}

Figure~\ref{fig:trapped_models_comparison} shows orbit-averaged integral spectra of trapped electrons and protons averaged over 60 days of orbiting (including SAA passages) with sampling of 10\,s obtained for different models by the AFRL package. It demonstrates that AE9/AP9 models gives much higher electron and proton fluxes compared to the AE8/AP8 models at low inclinations and low energies. Therefore the detected background count rate due to the trapped particles calculated by our simulation may overestimate the real level.

\begin{figure*}[h]
	\centerline{\includegraphics[width=0.99\linewidth]{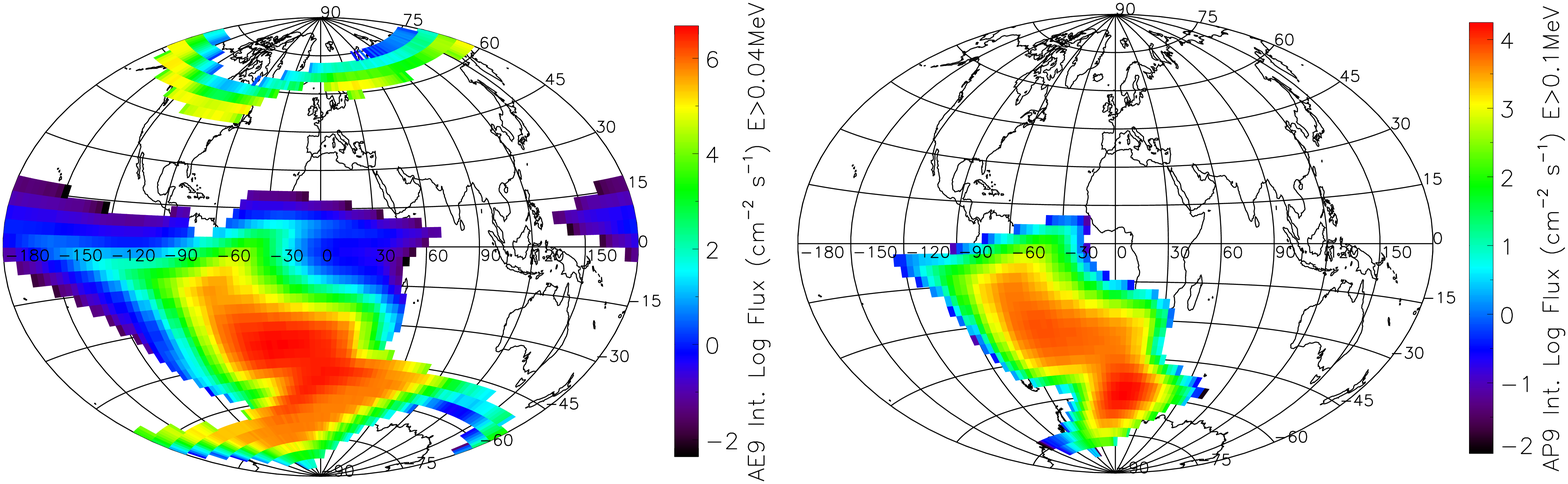}
	}
	\caption{A map of the integral flux of geomagnetically trapped electrons (left) and protons (right) at 500\,km altitude according to the AE9 and AP9 models (MC mode, 50\,\% CL), respectively, obtained by the AFRL package.\label{fig:AE9/AP9-maps}}
\end{figure*}

Figure~\ref{fig:AE9/AP9-maps} shows maps of integral fluxes (flux of particles with energy higher than $E$) of trapped electrons and protons for the AE9 and AP9 models with Monte Carlo (MC) mode, 100 runs and 50\,\% confidence level (CL) at 500\,km altitude. The MC mode accounts for the uncertainty due to the random perturbations as well as the flux variations due to the space weather \cite{Ginet2013}.

We calculated spectra of electrons and protons averaged along the trajectory of a satellite at altitude 500\,km with inclination of $i=20^\circ$ and orbiting 30 days with flux sampling every 10\,s. Only the regions with the integral flux $\leq10$\,particle\,cm$^{-2}$s$^{-1}$ were used ($E>40$\,keV for electrons and $E>100$\,keV for protons). These conditions give a duty cycle, i.e. the fraction of time a satellite spends in a region with particle flux lower than a given flux threshold, of 80\,\%. If the orbital inclination is $90^\circ$ the duty cycle would be 76\,\%. Details about the duty cycle for different inclinations, altitudes at LEO, flux and energy thresholds for AE8, AP8, AE9 and AP9 models can be found in Ref.~\citenum{Ripa2020}.

\begin{figure*}[h]
	\centerline{
	\includegraphics[width=0.99\linewidth]{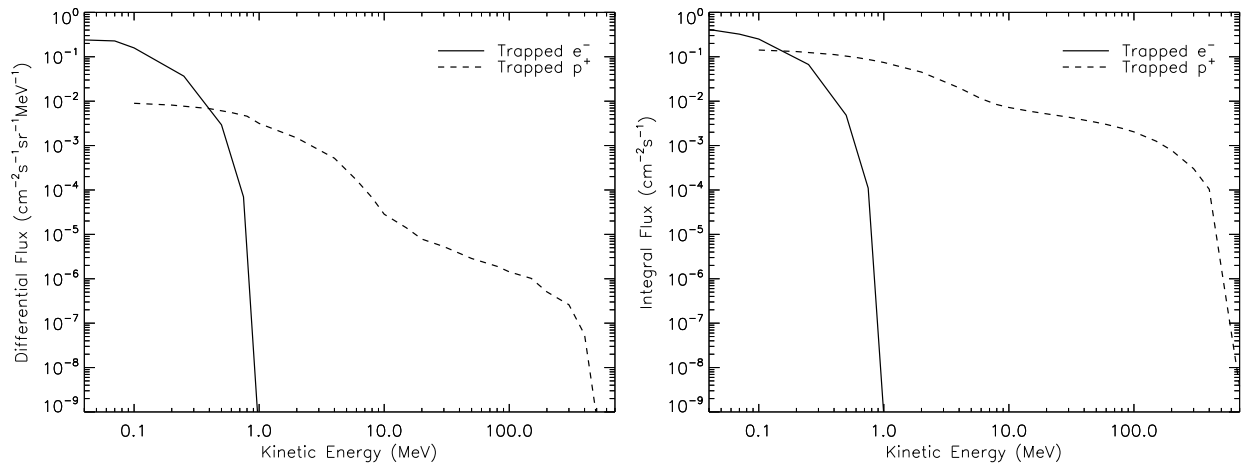}
	}
	\caption{Differential fluxes (left) and integral fluxes (right) of geomagnetically trapped electrons and protons averaged along the trajectory of a satellite at altitude 500\,km with inclination $i=20^\circ$ and orbiting 30 days. The models were AE9 and AP9, MC mode with 100 runs and the spectra were derived from the 50\,\% CL of the fluxes. Only the regions avoiding SAA were used. The differential flux per solid angle has been calculated for simplicity assuming the radiation is illuminating the satellite isotropically from the solid angle of $4\pi$.\label{fig:AE9/AP9-spectra}}
\end{figure*}

In this way we were able to obtain averaged spectra outside of SAA. The differential and integral fluxes are shown in Figure~\ref{fig:AE9/AP9-spectra}. The differential flux per solid angle has been calculated for simplicity assuming the radiation is illuminating a satellite isotropically from the solid angle of $4\pi$ because, for example, in case of \textit{CAMELOT} satellites the pointing strategy is not established yet. The trapped particles can collide with the detector from various directions and we are interested in a long-term average. Also, the assumption of the isotropy of the trapped particle flux is a simplification because of the well known ``East-West'' effect \cite{Ginet2007}. The integral flux for trapped electrons is 0.41\,cm$^{-2}$s$^{-1}$ ($E>40$\,keV) and for trapped protons is 0.14\,cm$^{-2}$s$^{-1}$ ($E>100$\,keV) as obtained from the model.

\subsection{Primary Cosmic-Rays}\label{sec:gcr}
The spectra of the primary particles of cosmic-rays (CRs) used in our simulations are described below. For the assumed 500\,km altitude the fluxes irradiate the satellite from the solid angle of 8.64\,sr. The origin of CRs is extraterrestrial consisting mainly of protons. Other components of CRs such as electrons, positrons, alpha particles and nuclei of heavier elements have been detected as well. Several experiments have been performed to study CRs, e.g. AMS \cite{Alcaraz2000a,Alcaraz2000b}, BESS \cite{Sanuki2000}, CREAM \cite{Yoon2011}, \textit{Fermi}/LAT \cite{Ackermann2010,Ackermann2014}, HESS \cite{Aharonian2008}, PAMELA \cite{Martucci2018}.

We considered two models for the spectra of primary particles. The first one was the ISO-15390\cite{ISO-15390} model, which is the international standard for estimating the radiation impact of CRs on hardware in space and which describes the fluxes of primary protons, alpha particles, and nuclei of heavier elements.

The second model which we considered was described by Mizuno et al. (2004), see Ref.~\citenum{Mizuno2004}, and it was based on measurements done by BESS and AMS experiments (see also Ref.~\citenum{Campana2013}). The flux of primary CRs in interstellar space can be modeled by a power law function:
\begin{equation}\label{eq:prim_part_intestellar}
F_\mathrm{U}(E_\mathrm{k}) = A\left[ \frac{R(E_\mathrm{k})}{\mathrm{GV}} \right]^{-a},
\end{equation}
where $R=pc/Ze$ is the rigidity of the particle as a function of its kinetic energy $E_\mathrm{k}$ or momentum $p$ and its charge $Ze$. The flux constant $A$ and the exponent $a$ are determined by fitting of the following model to the measurements.

The flux of primary CRs for a given phase of the solar cycle and in a given position in the Earth's magnetosphere according to the model described in Ref.~\citenum{Mizuno2004} is:
\begin{equation}\label{eq:prim_part_flux}
F(E_\mathrm{k}) = F_\mathrm{U}(E_\mathrm{k} + Ze\phi) \times F_\mathrm{M}(E_\mathrm{k}, M, Z, \phi) \times F_\mathrm{C}(R, h, \theta_\mathrm{M}),
\end{equation}
where $M$ is the mass of the particle, $\phi$ is a solar modulation potential, $h$ is the altitude of the satellite's orbit and $\theta_\mathrm{M}$ is the geomagnetic latitude.

By applying an effective shift of energy of the primary particles due to the deceleration by the solar wind the first function $F_\mathrm{U}$ in Eq.~(\ref{eq:prim_part_flux}) gets form:
\begin{equation}
F_\mathrm{U}(E_\mathrm{k} + Ze\phi) = A\left[ \frac{R(E_\mathrm{k} + Ze\phi)}{\mathrm{GV}} \right]^{-a}.
\end{equation}

The second function $F_\mathrm{M}$ accounts for the flux modulation due to the solar cycle and is given by:
\begin{equation}
F_\mathrm{M}(E_\mathrm{k}, M, Z, \phi) = \frac{(E_\mathrm{k}+Mc^2)^2 - (Mc^2)^2}{(E_\mathrm{k} + Ze\phi + Mc^2)^2 - (Mc^2)^2}\,,
\end{equation}
where the solar modulation potential varies between $\phi=0.55$\,GV for solar minimum and $\phi=1.10$\,GV for solar maximum.

The third term is the geomagnetic cutoff function $F_\mathrm{C}$ given by:
\begin{equation}
F_\mathrm{C}(R, h, \theta_\mathrm{M}) = \frac{1}{1 + (R/R_\mathrm{cut})^{-r}}\,,
\end{equation}
where $r=12$ for $p^+$ or $\alpha$ particles, $r=6$ for $e^-$ or $e^+$, and the cutoff rigidity $R_\mathrm{cut}$ is given by the St\"ormer equation \cite{smart2005}:
\begin{equation}\label{eq:cutoff_rigidity}
R_\mathrm{cut} = 14.5\left(1 + \frac{h}{R_\mathrm{E}}\right)^{-2}\cos^4\theta_\mathrm{M}\,\mathrm{GV},
\end{equation}
where $R_\mathrm{E}$ is the Earth's radius.

We want to estimate a long term average of the flux, therefore we assume that the flux has uniform angular distribution for the zenith angle $0^\circ \leq \theta \leq \theta_\mathrm{cut}$ and the flux is zero for $\theta_\mathrm{cut} \leq \theta \leq 180^\circ$, where the $\theta_\mathrm{cut}$ is the zenith angle of the Earth's horizon and it is $112^\circ$ for the altitude of 500\,km.

\subsubsection{Primary Protons and Alpha Particles}\label{sec:gcr_H_He}

For primary protons and alpha particles we compared the model ISO-15390 and the model Mizuno et al. (2004) described in Ref.~\citenum{Mizuno2004}.

For the ISO-15390 model we employed SPENVIS where we generated the orbit-averaged spectra for circular orbit with inclination $i=20^\circ$, duration 30 days and sampling 60\,s. The following parameters setting was applied: solar minimum activity (May 1996), magnetic shielding on, stormy and quite magnetosphere, St\o{}rmer with eccentric dipole method and magnetic field moment unchanged.

For the model described in Ref.~\citenum{Mizuno2004} we used the following parameters: solar minimum cycle with the solar modulation potential $\phi=0.55$\,GV, altitude 500\,km, and two geomagnetic latitudes $\theta_\mathrm{M}=0^\circ$ and $\theta_\mathrm{M}=20^\circ$ (orbital inclination) + $9.6^\circ$ (tilt between the geomagnetic dipole axis and the Earth's rotational axis).

For primary protons the values of $A=23.9$\,particle\,m$^{-2}$s$^{-1}$sr$^{-1}$MeV$^{-1}$ and $a=2.83$ were adopted. For primary alpha particles the values of $A=1.5$\,particle\,m$^{-2}$s$^{-1}$sr$^{-1}$MeV$^{-1}$ and $a=2.77$ were adopted.

\begin{figure*}[h]
	\centerline{
	\includegraphics[width=0.99\linewidth]{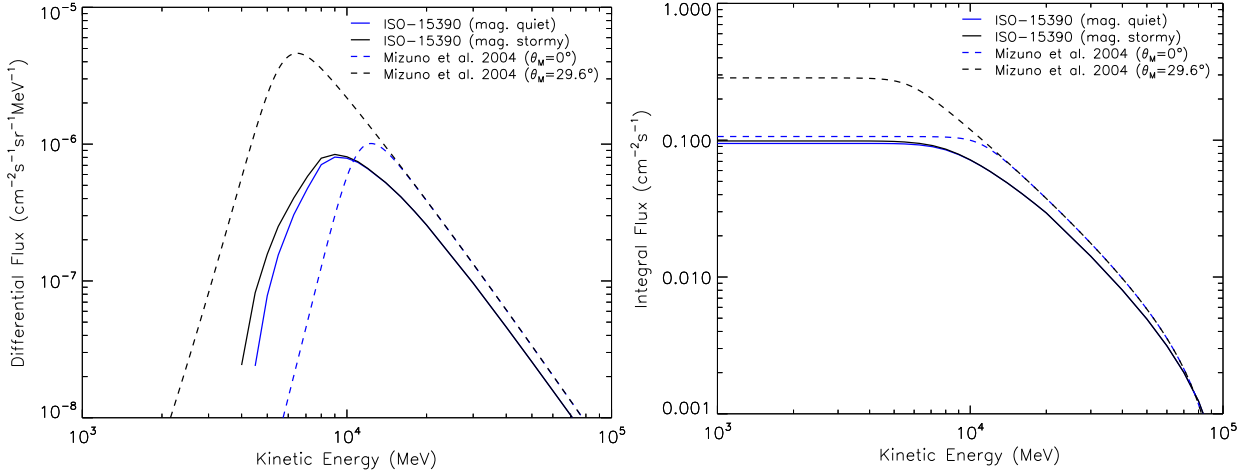}\
	}
	\caption{Differential fluxes (left) and integral fluxes (right) of primary CR protons. Solid lines mark model introduced by Mizuno et al. (2004) \cite{Mizuno2004} for geomagnetic latitudes $\theta_\mathrm{M}=0^\circ$ and $\theta_\mathrm{M}=29.6^\circ$. Dashed lines mark galactic CR model ISO-15390 \cite{ISO-15390} for quiet and stormy magnetosphere obtained in SPENVIS for circular orbit with inclination $i=20^\circ$. Spectra from both models were obtained for altitude of 500\,km. The integral flux is multiplied by a solid angle corresponding to the radiation illuminating the satellite at this altitude.
\label{fig:primary_protons}}
\end{figure*}

\begin{figure*}[h]
	\centerline{
	\includegraphics[width=0.99\linewidth]{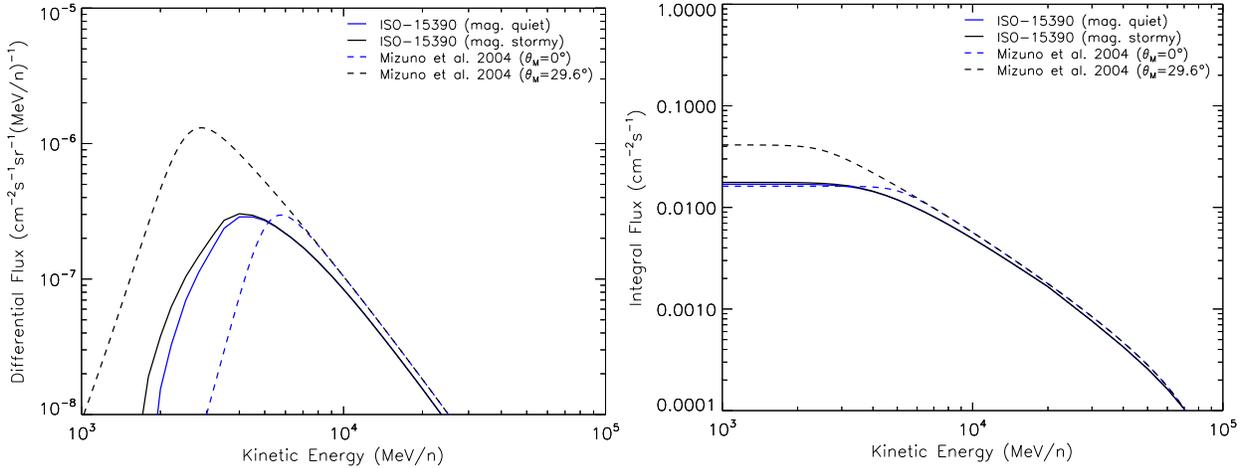}
	}
	\caption{Differential fluxes (left) and integral fluxes (right) of primary CR alpha particles as a function of energy per nucleon. Solid lines mark model introduced by Mizuno et al. (2004) \cite{Mizuno2004} for geomagnetic latitudes $\theta_\mathrm{M}=0^\circ$ and $\theta_\mathrm{M}=29.6^\circ$. Dashed lines mark galactic CR model ISO-15390 \cite{ISO-15390} for quiet and stormy magnetosphere obtained in SPENVIS for circular orbit with inclination $i=20^\circ$. Spectra from both models were obtained for altitude of 500 km. The integral flux is multiplied by a solid angle corresponding to the radiation illuminating the satellite at this altitude.
\label{fig:primary_alpha}}
\end{figure*}

Figure~\ref{fig:primary_protons} and \ref{fig:primary_alpha} show fluxes of primary protons and alpha particles, respectively, at 500\,km altitude obtained by the ISO-15390 model for quiet and stormy magnetosphere and obtained by the Mizuno et al. (2004) model for fixed geomagnetic latitudes $\theta_\mathrm{M}=0^\circ$ and $\theta_\mathrm{M}=29.6^\circ$. This model predicts the integral flux of primary protons of kinetic energies $E>1$\,GeV being 0.11\,cm$^{-2}$s$^{-1}$ (for $\theta_\mathrm{M}=0^\circ$) or 0.29\,cm$^{-2}$s$^{-1}$ (for $\theta_\mathrm{M}=29.6^\circ$) and the integral flux of primary alpha particles of kinetic energies per nucleon $E>1$\,GeV/n being 0.016\,cm$^{-2}$s$^{-1}$ (for $\theta_\mathrm{M}=0^\circ$) or 0.041\,cm$^{-2}$s$^{-1}$ ($\theta_\mathrm{M}=29.6^\circ$).

For our Geant4 simulations of the expected on-board background by a \textit{CAMELOT} satellite (Sec.~\ref{sec:sim_results}) we use the ISO-15390 model with stormy magnetosphere, because it is the international standard for CR flux and because it was obtained for a fixed orbital inclination of $i=20^\circ$ meaning crossing the range of geomagnetic latitudes between $-30^\circ$ and $\sim32^\circ$\footnote{\url{https://spawx.nwra.com/spawx/maps/maplats.html}}$^,$ \cite{VanZandt1972}. It predicts the integral flux of primary protons of kinetic energies $E>1$\,GeV being 0.095\,cm$^{-2}$s$^{-1}$ (for quite magnetosphere) or 0.099\,cm$^{-2}$s$^{-1}$ (for stormy magnetosphere) and the integral flux of primary alpha particles of kinetic energies per nucleon $E>1$\,GeV/n being 0.017\,cm$^{-2}$s$^{-1}$ (for quite magnetosphere) or 0.018\,cm$^{-2}$s$^{-1}$ (for stormy magnetosphere).

\subsubsection{Primary Electrons and Positrons}\label{sec:gcr_el_pos}

For the spectra of the primary e$^-$ and e$^+$ we used the model described by Mizuno et al. (2004) \cite{Mizuno2004} with references to the flux measurements by \cite{Webber1983,Moskalenko1998}. The measurements of the ratio of positrons and electrons e$^+$/(e$^-$+e$^+$) are given by \cite{Golden1994,Alcaraz2000b}.

We adopt the model of the primary interstellar particles Eq.~(\ref{eq:prim_part_intestellar}) with following parameters: $A=0.65$\,particle\,m$^{-2}$s$^{-1}$sr$^{-1}$MeV$^{-1}$ and $a=3.3$ for electrons and $A=0.051$\,particle\,m$^{-2}$s$^{-1}$sr$^{-1}$MeV$^{-1}$ with the same exponent $a$ for positrons.

\begin{figure*}[h]
	\centerline{
	\includegraphics[width=0.99\linewidth]{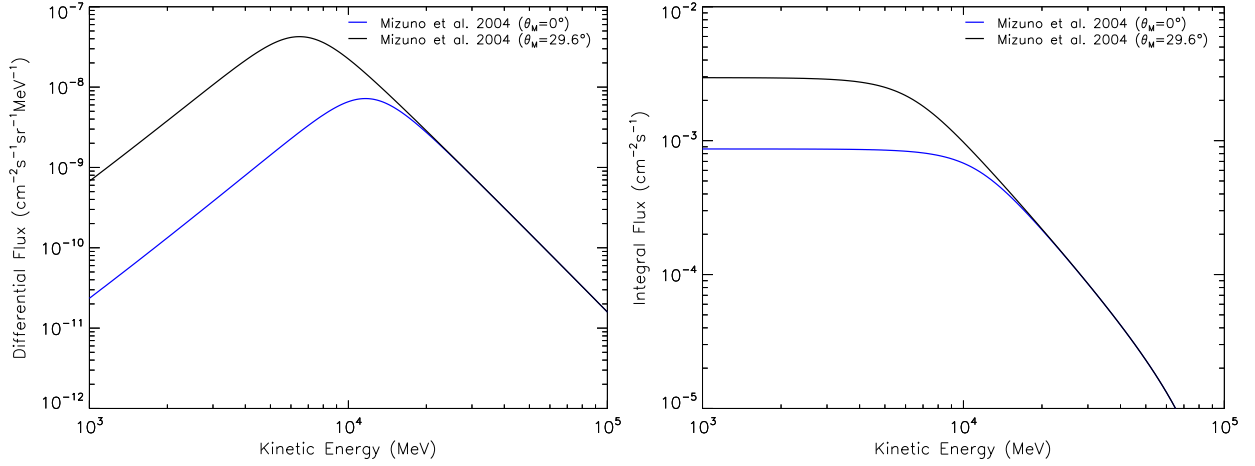}
	}
	\caption{Differential fluxes (left) and integral fluxes (right) of primary CR electrons modeled by Mizuno et al. (2004) \cite{Mizuno2004} for geomagnetic latitudes $\theta_\mathrm{M}=0^\circ$ and $\theta_\mathrm{M}=29.6^\circ$ and altitude of 500\,km. The integral flux is multiplied by a solid angle corresponding to the radiation illuminating the satellite at this altitude.
\label{fig:primary_electrons}}
\end{figure*}

\begin{figure*}[h]
	\centerline{
	\includegraphics[width=0.99\linewidth]{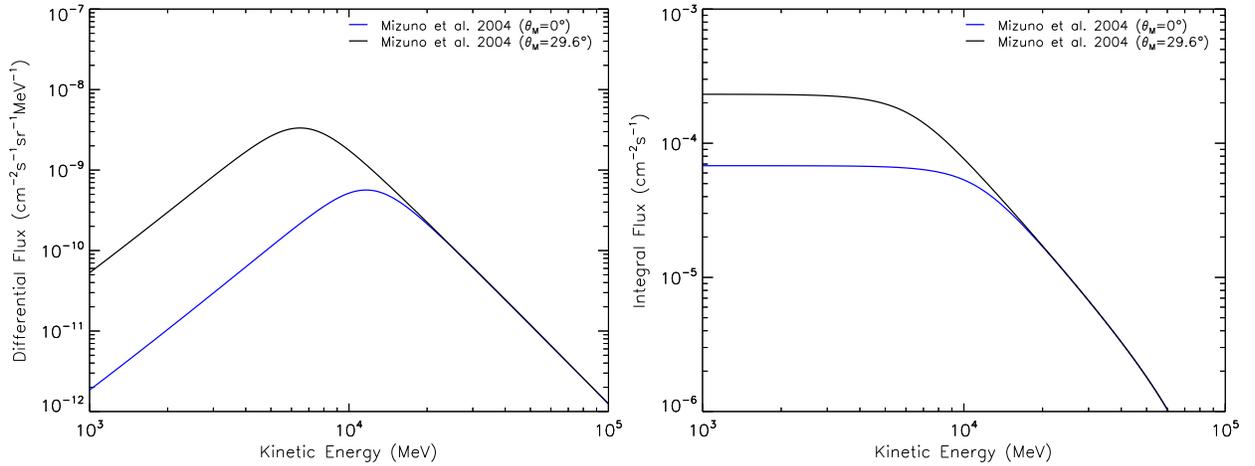}
	}
	\caption{Differential fluxes (left) and integral fluxes (right) of primary CR positrons modeled by Mizuno et al. (2004) \cite{Mizuno2004} for geomagnetic latitudes $\theta_\mathrm{M}=0^\circ$ and $\theta_\mathrm{M}=29.6^\circ$ and altitude of 500\,km. The integral flux is multiplied by a solid angle corresponding to the radiation illuminating the satellite at this altitude.
\label{fig:primary_positrons}}
\end{figure*}

Same as for the primary protons and alpha particles we used the following conditions of the solar cycle and the orbit: solar minimum with the modulation potential $\phi=0.55$\,GV, altitude 500\,km, and two geomagnetic latitudes $\theta_\mathrm{M}=0^\circ$ and $\theta_\mathrm{M}=29.6^\circ$. Figure~\ref{fig:primary_electrons} and \ref{fig:primary_positrons} show the fluxes of the primary electrons and positrons.

For our simulations of the expected detected background (Sec.~\ref{sec:sim_results}) we use the spectrum for $\theta_\mathrm{M}=29.6^\circ$. We also assume that the angular distribution of the flux is incoming uniformly from the solid angle unocculted by the Earth. Then the integral flux ($E>1$\,GeV) for e$^-$ is $3\times10^{-3}$\,cm$^{-2}$s$^{-1}$ and for e$^+$ it is $2.3\times10^{-4}$\,cm$^{-2}$s$^{-1}$.

\subsection{Secondary Particles and Radiation}\label{sec:secondary_rad}
Secondary (albedo) particles and radiation are created by interaction of primary CRs with the Earth's atmosphere \cite{Jursa1985,Zombeck2007}.

\subsubsection{Secondary Protons}\label{sec:secondary_p}
For secondary p$^+$ and for energy above 100\,MeV we use the model \cite{Mizuno2004} based on the measurements done by the Alpha Magnetic Spectrometer (AMS) \cite{Alcaraz2000a} from 380\,km altitude for the geomagnetic latitude $0.3 \textrm{ rad} \leq \theta_\textrm{M} \leq 0.4 \textrm{ rad}$. For energy below 100\,MeV we use the fit to \textit{MITA}/NINA-2 data \cite{Bidoli2002} from 450\,km altitude and for $1.0 \leq \textrm{L-shell} \leq 1.7$, where L-shell is the McIlwain L-parameter \cite{McIlwain1961}. For details see the LAT Technical Note LAT-TD-08316-01 \cite{LAT-TD-08316-01} of the \textit{Fermi} satellite \cite{Atwood2009}. There is only small dependence of the flux on altitude \cite{Bidoli2002,Zuccon2003} therefore it can be used as an approximation to the flux at altitude of 500\,km.

\begin{figure*}[h]
	\centerline{
	\includegraphics[width=0.99\linewidth]{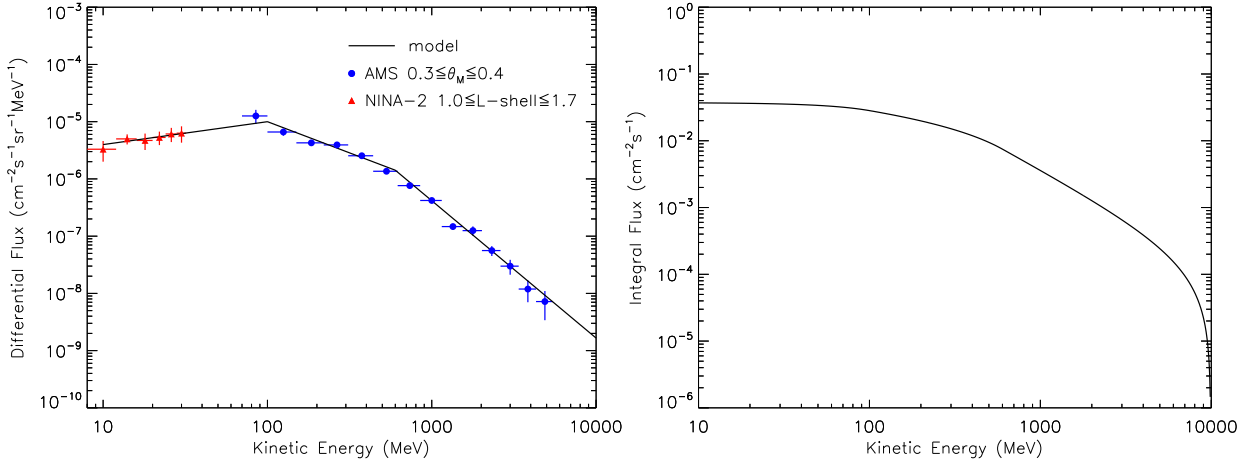}
	}
	\caption{Differential fluxes (left) and integral fluxes (right) of secondary protons modeled by Eq.~(\ref{eq:secondary_protons}) are marked by the black curve. The measurements from the AMS and \textit{MITA}/NINA-2 experiments for the given geomagnetic position are shown as well. The integral flux is multiplied by the solid angle of $4\pi$\,sr.
\label{fig:secondary_protons}}
\end{figure*}

The differential flux $F(E)$ in units of particle\,m$^{-2}$s$^{-1}$sr$^{-1}$MeV$^{-1}$ is modeled as:
 \begin{equation}\label{eq:secondary_protons}
 F(E)=
 \begin{cases}
 0.1\left ( \frac{E}{100\,\text{MeV}} \right )^{0.4} & \text{for } 10\,\text{MeV} \leq E \leq 100\,\text{MeV} \\
  0.1\left ( \frac{E}{100\,\text{MeV}} \right )^{-1.09} & \text{for } 100\,\text{MeV} \leq E \leq E_\mathrm{brk} \\
 0.1\left ( \frac{E_\mathrm{brk}}{100\,\text{MeV}} \right )^{-1.09} \left ( \frac{E}{E_\mathrm{brk}} \right )^{-2.4} & \text{for } E \geq E_{\mathrm{brk}},
 \end{cases}
 \end{equation}
where $E_\mathrm{brk}=600$\,MeV is break energy. Figure~\ref{fig:secondary_protons} shows the modeled flux together with the measurements.

The same model is used for the upward and downward component of the flux and it is assumed that secondary protons irradiate the satellite from the solid angle of $4\pi$\,sr without zenith-angle dependence of the flux. The integral flux ($E\geq10$\,MeV) is 0.037\,cm$^{-2}$s$^{-1}$.

\subsubsection{Secondary Electrons and Positrons}\label{sec:secondary_el_pos}
For secondary e$^-$ and e$^+$ and for energy above 100\,MeV we use the model \cite{Mizuno2004} based on the measurements done by AMS \cite{Alcaraz2000a} from 380\,km altitude for the geomagnetic latitude $0 \leq \theta_\textrm{M} \leq 0.3 \textrm{ rad}$. For energy below 100\,MeV we use the fit to \textit{Mir}/MARIA-2 data \cite{Voronov1991,Mikhailov2002} from 400\,km altitude and for $1.0 \leq \textrm{L-shell} \leq 1.2$. For details see the LAT Technical Note LAT-TD-08316-01 \cite{LAT-TD-08316-01}.

\begin{figure*}[h]
	\centerline{
	\includegraphics[width=0.99\linewidth]{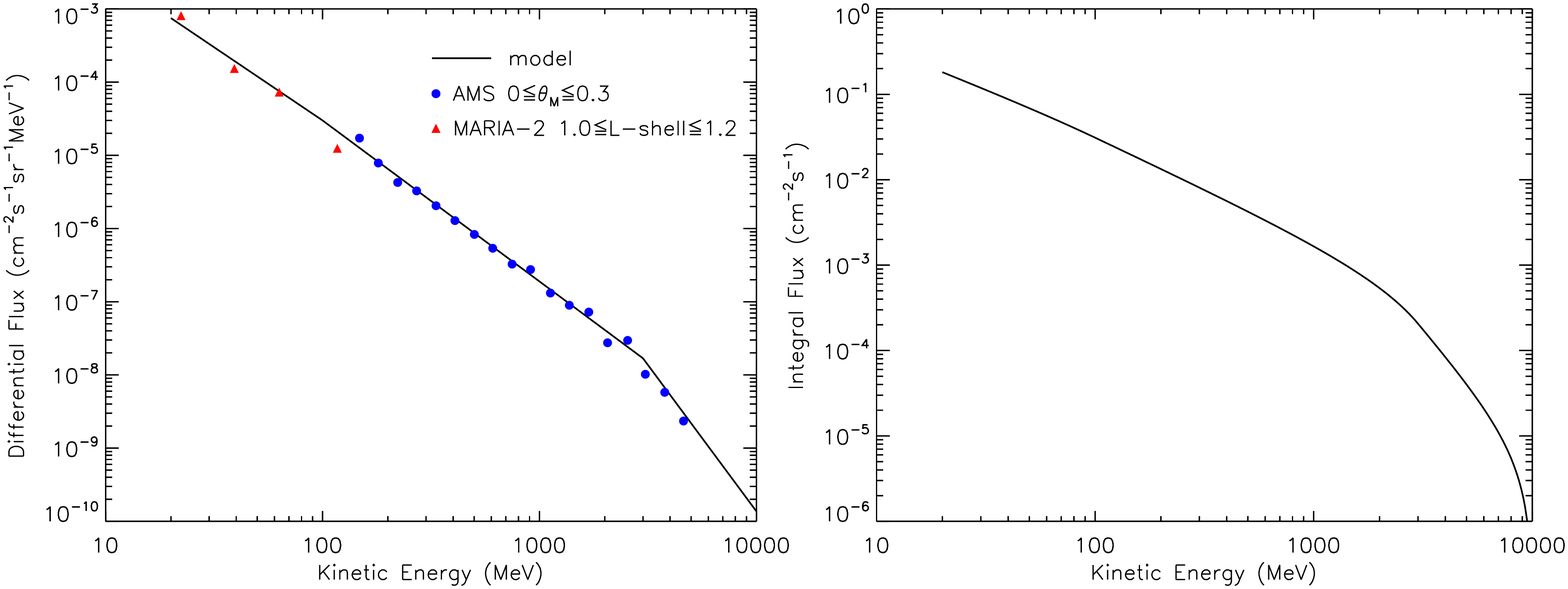}
	}
	\caption{Differential fluxes (left) and integral fluxes (right) of secondary electrons modeled by Eq.~(\ref{eq:secondary_electrons}) are marked by the black curve. The measurements from the AMS and \textit{Mir}/MARIA-2 experiments for the given geomagnetic position are shown as well. The integral flux is multiplied by the solid angle of $4\pi$\,sr.
\label{fig:secondary_electrons}}
\end{figure*}

For secondary e$^-$ the differential flux $F(E)$ in units of particle\,m$^{-2}$s$^{-1}$sr$^{-1}$MeV$^{-1}$ is modeled as:
 \begin{equation}\label{eq:secondary_electrons}
 F(E)=
 \begin{cases}
 0.3\left ( \frac{E}{100\,\text{MeV}} \right )^{-2.0} & \text{for } 10\,\text{MeV} \leq E \leq 100\,\text{MeV} \\
 0.3\left ( \frac{E}{100\,\text{MeV}} \right )^{-2.2} & \text{for } 100\,\text{MeV} \leq E \leq E_\mathrm{brk} \\
 0.3\left ( \frac{E_\mathrm{brk}}{100\,\text{MeV}} \right )^{-2.2} \left ( \frac{E}{E_\mathrm{brk}} \right )^{-4.0} & \text{for } E \geq E_{\mathrm{brk}},
 \end{cases}
 \end{equation}
where the break energy $E_\mathrm{brk}=3000$\,MeV. Figure~\ref{fig:secondary_electrons} shows the modeled flux together with the measurements.

\begin{figure*}[h]
	\centerline{
	\includegraphics[width=0.99\linewidth]{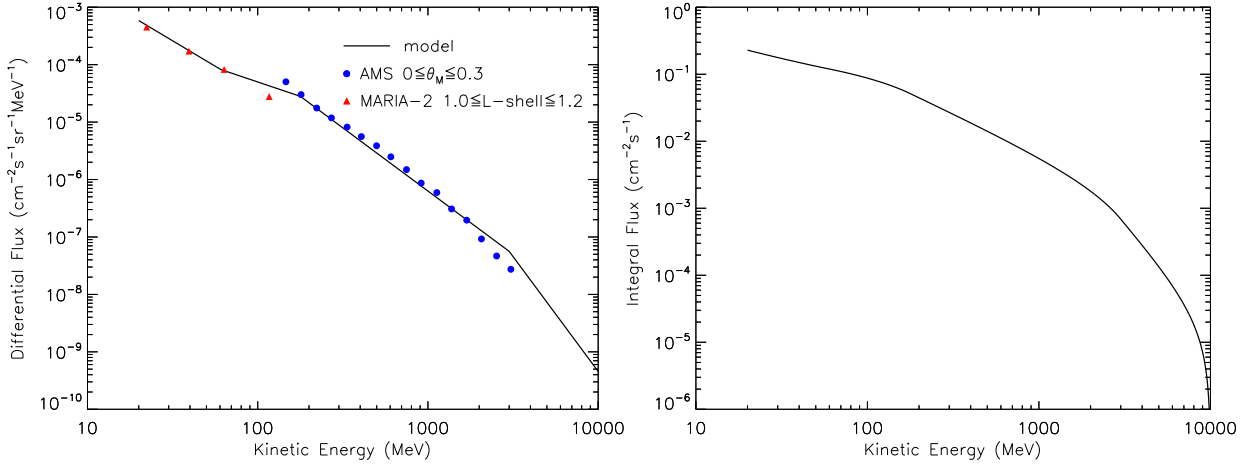}
	}
	\caption{Differential fluxes (left) and integral fluxes (right) of secondary positrons modeled by Eq.~(\ref{eq:secondary_positrons}) are marked by the black curve. The measurements from the AMS and \textit{Mir}/MARIA-2 experiments for the given geomagnetic position are shown as well. The integral flux is multiplied by the solid angle of $4\pi$\,sr.
\label{fig:secondary_positrons}}
\end{figure*}

For secondary e$^+$ the differential flux $F(E)$ in units of particle\,m$^{-2}$s$^{-1}$sr$^{-1}$MeV$^{-1}$ is modeled as:
 \begin{equation}\label{eq:secondary_positrons}
 F(E)=
 \begin{cases}
 20\left ( \frac{E}{10\,\text{MeV}} \right )^{-1.77} & \text{for }  10\,\text{MeV} \leq E \leq 60\,\text{MeV} \\
 0.833\left ( \frac{E}{60\,\text{MeV}} \right )^{-1.0} & \text{for } 60\,\text{MeV} \leq E \leq 178\,\text{MeV} \\
 1.0\left ( \frac{E}{100\,\text{MeV}} \right )^{-2.2} & \text{for } 178\,\text{MeV} \leq E \leq E_\mathrm{brk} \\
 1.0\left ( \frac{E_\mathrm{brk}}{100\,\text{MeV}} \right )^{-2.2} \left ( \frac{E}{E_\mathrm{brk}} \right )^{-4.0} & \text{for } E \geq E_{\mathrm{brk}},
 \end{cases}
 \end{equation}
where the break energy $E_\mathrm{brk}=3000$\,MeV. Figure~\ref{fig:secondary_positrons} shows the modeled flux together with the measurements.

The same model is used for the upward and downward component of the flux and it is assumed that secondary e$^-$ and e$^+$ irradiate the satellite from the solid angle of $4\pi$\,sr without zenith-angle dependence of the flux. The integral flux ($E\geq20$\,MeV) is 0.18\,cm$^{-2}$s$^{-1}$ for e$^-$ and 0.23\,cm$^{-2}$s$^{-1}$ for e$^+$.

\subsubsection{Albedo X-rays/$\gamma$-rays}\label{sec:albedo_x-rays}
The secondary (albedo) X-ray and $\gamma$-ray flux is due to interaction of primary CRs with the Earth's atmosphere. It is produced by decay of $\pi^0$ pions (mainly above 50\,MeV), by bremsstrahlung from primary and secondary electrons (mainly below 50\,MeV), and also by the reflection of CXB and it has been measured by several satellites and balloon experiments \cite{Schwartz1974, Thompson1974, Schoenfelder1977, Schoenfelder1980, Gurian1979, Ryan1979, Gehrels1992, Akyuz1997, Dean2003, Mizuno2004, Petry2005, Zombeck2007, Churazov2007, Abdo2009, Turler2010, Cumani2019}. The intensity depends on the geomagnetic latitude \cite{Imhof1976}.

We utilize a model reported by Ajello et al. (2008) \cite{Ajello2008} based on the \textit{Swift}/BAT measurements from $\sim20$\,keV to $\sim200$\,keV for altitude of $h\sim550$\,km and inclination of $i=20.6^\circ$ and which is compatible with measurements from BeppoSAX\cite{Frontera2007} ($h\sim580$\,km and $i=4^\circ$) and after some corrections with measurements by the polar-orbiting satellite 1972-076B \cite{Imhof1976} ($h\sim750$\,km). MC simulations show that this model is a very good approximation of the Earth albedo X-ray emission up to 300\,keV \cite{Sazonov2007, Ajello2008}.

We assume the Ajello et al. (2008) model \cite{Ajello2008} in the energy range of $E=10-300$\,keV and hence the differential photon flux $F(E)$ given by Eq.~(\ref{eq:Ajello2008}), where the model parameters and their 90\,\% CL errors are $\Gamma_1$=-5 (fixed), $\Gamma_2$=1.72$\pm0.08$, $E_\mathrm{b}$=33.7$\pm3.5$\,keV and $C=1.48^{+0.6}_{-0.3}\times10^{-2}$.

\begin{figure*}[h]
	\centerline{
	\includegraphics[width=0.99\linewidth]{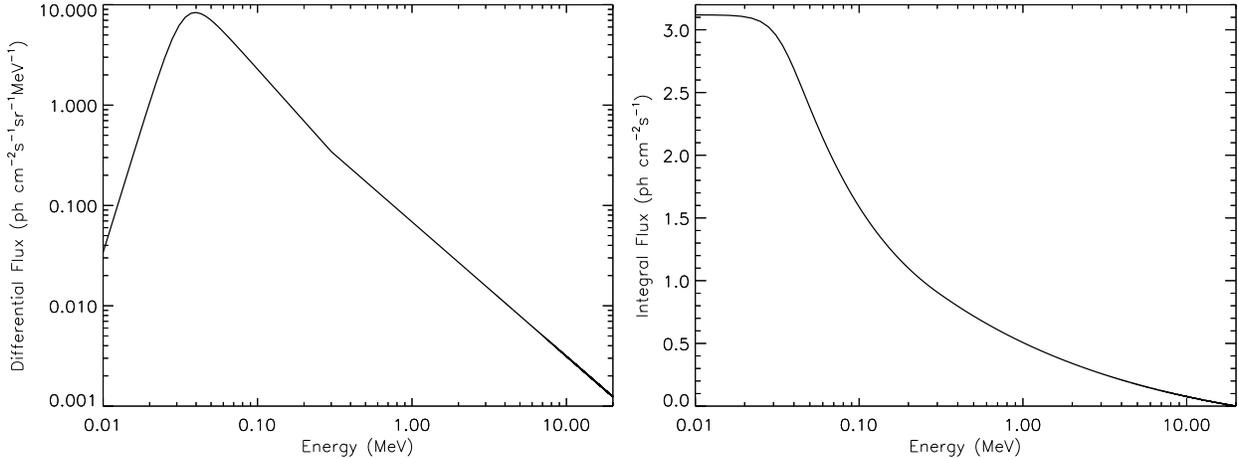}
	}
	\caption{The albedo X-ray/$\gamma$-ray spectra. For energies from 10\,keV to 300\,keV modeled by Ajello et al. (2008) and for energies from 300\,keV to 20\,MeV modeled by Eq.~(\ref{eq:Mizuno2004_albedo_gamma}). \textit{Left:} The differential photon flux. \textit{Right:} The integral photon flux integrated up to 20\,MeV. The integral flux is multiplied by the Earth-subtended solid angle of 3.93\,sr at an altitude of 500 km. \label{fig:albedo_x-rays}}
\end{figure*}

For higher energies we assume a model reported by Mizuno et al. (2004) \cite{Mizuno2004} based on measurements by \textit{1972-076B} and \textit{Kosmos 461} satellites \cite{Imhof1976, Gurian1979} and by balloon flights \cite{Thompson1974, Ryan1979}. Particularly, we consider only energies $E=0.3-20$\,MeV where we assume the differential photon flux $F(E)$ in units of ph\,cm$^{-2}$s$^{-1}$sr$^{-1}$keV$^{-1}$ to be a simple power law function:
\begin{equation}\label{eq:Mizuno2004_albedo_gamma}
F(E) = 719\left ( \frac{E}{\text{keV}} \right )^{-1.34},
\end{equation}
where we normalized the Mizuno et al. (2004) model \cite{Mizuno2004}, their Eq.~(21), in order to obtain the same differential flux at 300\,keV as predicted by the Ajello et al. (2008) model \cite{Ajello2008}. The spectrum is shown in Figure~\ref{fig:albedo_x-rays}. According to MC simulations \cite{Sazonov2007} there is only a small dependence of the albedo X-ray flux ($25-300$\,keV) on the solar cycle and geomagnetic latitude below $\sim20^\circ$ for an instrument at LEO with large FOV which covers the whole terrestrial disc.

A zenith angle dependence of the albedo $\gamma$-ray flux has been measured in the $1-10$\,MeV region \cite{Schoenfelder1977, Schwartz1974, Mizuno2004}. See also Ref.~\citenum{Kraushaar1972, Thompson1974, Imhof1976, Ryan1979, Thompson1981, Akyuz1997, Dean2003, Petry2005, Abdo2009} and references therein for the zenith angle dependence of the albedo $\gamma$-ray flux at other energies. In the energy range $25-300$\,keV, covered by the Ajello et al. (2008) model, the MC simulations \cite{Sazonov2007} suggest that there is no zenith angle dependence. However, for the higher-energy part $0.3-20$\,MeV one can expect a zenith angle dependence of the flux. In case of \textit{CAMELOT} satellites, they will have detectors with all-sky FOV which can be illuminated from various directions and we are interested in a long term average flux, therefore, for simplicity, we do not assume any zenith angle dependence in our Geant4 simulations involving a \textit{CAMELOT} satellite mass model.

At an altitude of 500\,km the photons would irradiate the satellite from a solid angle of 3.93\,sr. The integral flux ($E>10$\,keV) is 3.1\,ph\,cm$^{-2}$s$^{-1}$.

\subsubsection{Albedo Neutrons}\label{sec:albedo_n}
The albedo neutrons are produced in hadronic showers created by CRs interacting with the Earth's atmosphere and they can reach a satellite at LEO \cite{Cumani2019}. For the albedo neutrons we use the predictions of the QinetiQ Atmospheric Radiation Model (QARM), based on MC radiation transport code, as reported in the ESA document ECSS-E-ST-10-04C \cite{ECSS-E-ST-10-04C}. The model has been validated against several measurements \cite{Ait-Ouamer1988,Morris1995,Lei2004,Lei2006} and is also consistent with other MC simulations\cite{Armstrong1992,Fioretti2012}. For other models and measurements see Ref.~\citenum{Lingenfelter1963,Dean2003,Kole2015,Cumani2019} and references therein.

\begin{figure*}[h]
	\centerline{
	\includegraphics[width=0.99\linewidth]{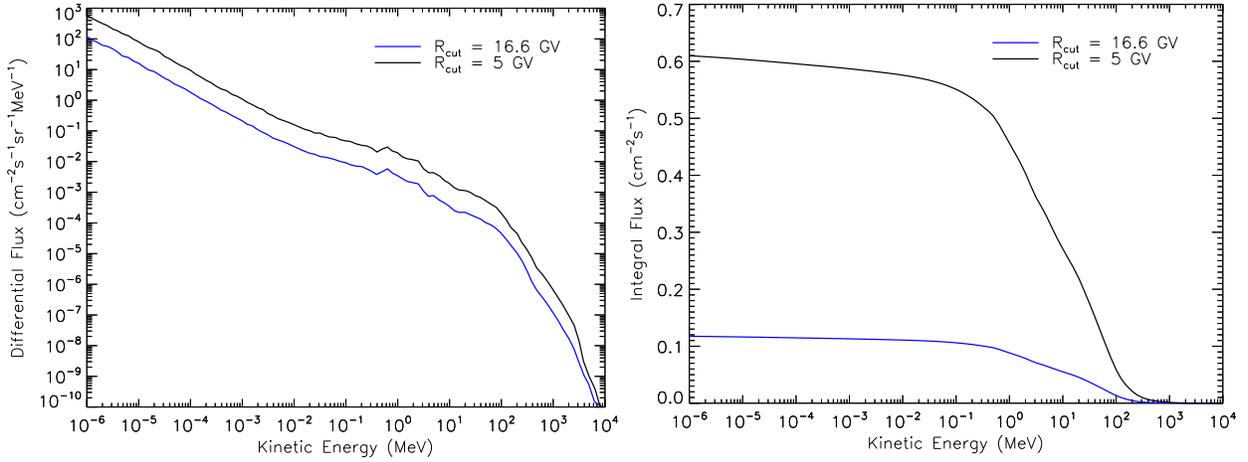}
	}
	\caption{Differential fluxes (left) and integral fluxes (right) of albedo neutrons predicted by the QARM model for two values of cutoff rigidity and scaled to the altitude of 500\,km. The integral flux is multiplied by the solid angle of 3.93\,sr.
\label{fig:albedo_n}}
\end{figure*}

Figure~\ref{fig:albedo_n} shows the fluxes of secondary neutrons for the cutoff rigidity $R_\mathrm{cut}=16.6$\,GV and $R_\mathrm{cut}=5$\,GV for solar minimum. The fluxes were scaled from the altitude of 100\,km to 500\,km as described in the ECSS-E-ST-10-04C document.

In our simulations of the expected detected background (Sec.~\ref{sec:sim_results}) we use the spectrum for the solar minimum and for the cutoff rigidity $R_\mathrm{cut}=5$\,GV which corresponds to the geomagnetic latitude $\theta_\mathrm{M}=37^\circ$ following from the St\"ormer equation Eq.~(\ref{eq:cutoff_rigidity}) or latitude between $\sim30^\circ$ and $\sim50^\circ$, see Figure~7 of Ref.~\citenum{smart2005}.

The integral flux is 0.61\,cm$^{-2}$s$^{-1}$ for $E>1$\,eV, for $R_\mathrm{cut}=5$\,GV, altitude of 500\,km and assuming that all neutrons are coming from the solid angle of 3.93\,sr which corresponds to the angular size of the Earth observed from that altitude.

\section{The \textit{CAMELOT} CubeSats}
\label{sec:camelot}

We study in particular the expected on-board background for the proposed \textit{CAMELOT} mission, expected to be launched to LEO with the main objective of all-sky monitoring and timing-based localization of GRBs. The at least nine satellites are considered to be placed on orbits with altitude of $\sim 500-600$\,km with inclination of $53^\circ$ or at Sun-synchronous orbits of inclination $97.6^\circ$ \cite{Werner2018}. One of the options for the \textit{CAMELOT} satellite platform is the one being developed by C3S LLC in Budapest, therefore we apply its mass model in our Geant4 simulations.

\subsection{The Detector System}

\begin{figure*}[h]
	\centerline{
	\includegraphics[width=0.99\linewidth]{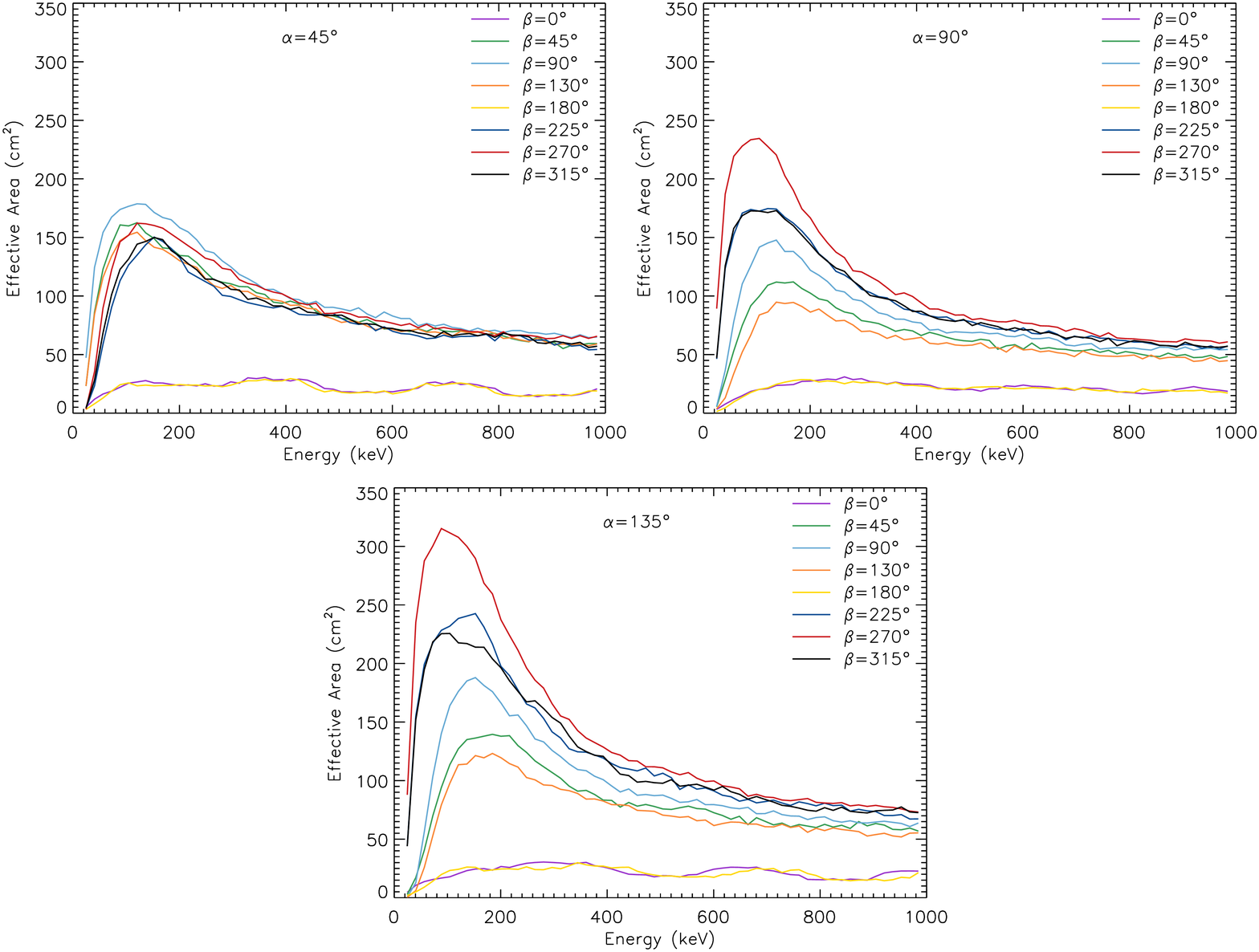}
	}
	\caption{The effective area of four detectors on board one \textit{CAMELOT} satellite as a function of energy and for different angles $\alpha$ and $\beta$ defining the source direction in respect to the satellite. For the exact definition of these angles see Fig.~\ref{fig:mass_model}.
\label{fig:eff_area}}
\end{figure*}

The constellation of at least nine 3U CubeSats is proposed to be equipped with large and thin CsI(Tl) scintillators, of size $75\times150\times5$~mm$^3$ each, read out by Hamamatsu Multi-Pixel Photon Counters (MPPC). There would most likely be four scintillators on each satellite with two scintillators placed on two neighbouring sides of the satellite. The scintillators will be wrapped in the enhanced specular reflector (ESR) foil and enclosed in a support structure made either from aluminium or carbon fiber-reinforced plastic (CFRP). For details about the detector system see Ref.~\citenum{Ohno2018}. The effective area of four detectors on board one \textit{CAMELOT} satellite as a function of energy and for different directions, obtained from Geant4 simulations, is shown in Figure~\ref{fig:eff_area}.

In order to understand and characterize the behaviour of the large-area CsI(Tl) scintillator detector and the MPPC readout, an experimental setup was built in Hiroshima, Japan. The experimental setup provided vital information for the simulation, mostly for the position dependence of the scintillator effective light yield. Different $\gamma$ sources were used in the tests, mostly an $^{241}$Am source \cite{Torigoe2019}.

\section{Validation of Geant4 Simulation and Calibration of Detector's Optical Parameters}
\label{sec:validate}

\begin{figure}[h]
\begin{center}
\begin{tabular}{c}
\includegraphics[width=0.7\linewidth]{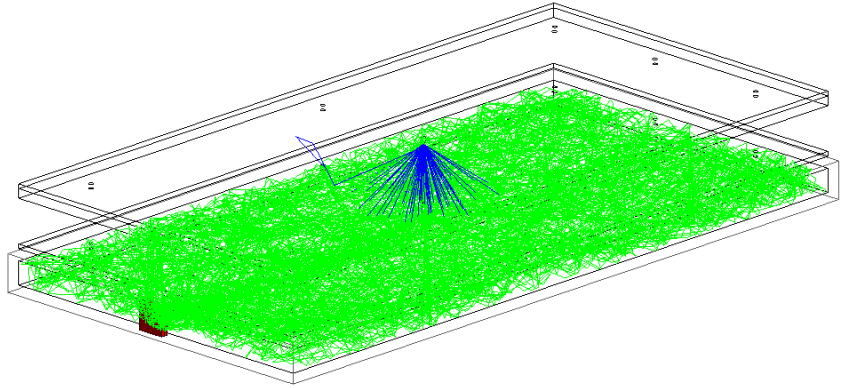}
\end{tabular}
\end{center}
\caption 
{ \label{fig:sim}
Simulation of 50 X-rays originating from a collimated X-ray source placed above the middle of the scintillator. Blue tracks are X-rays, green tracks are optical photons and the red square marks the MPPC. Only optical photons which are detected were drawn.} 
\end{figure}

 A dedicated set of measurements were carried out with $^{241}$Am $\gamma$ source with an activity of 471\,kBq which was collimated to irradiate different positions on the scintillator. The experiments were carried out with a single MPPC very similarly to the measurements presented in Ref.~\citenum{Torigoe2019}. The collimation was achieved with two lead sheets each containing holes in nine positions. In order to obtain the optical parameters of the scintillators as precisely as possible, the effect of reflectivity and absorption length on photon light yield was maximized by utilizing one MPPC in the middle of the shorter side of the scintillator. Spectra were recorded in nine cases by moving the $^{241}$Am source in the nine positions where holes were present. Figure~\ref{fig:sim} shows the simulation of the experimental set-up.
 
\begin{figure*}[h]
 	\centerline{
 	\includegraphics[width=1.0\linewidth]{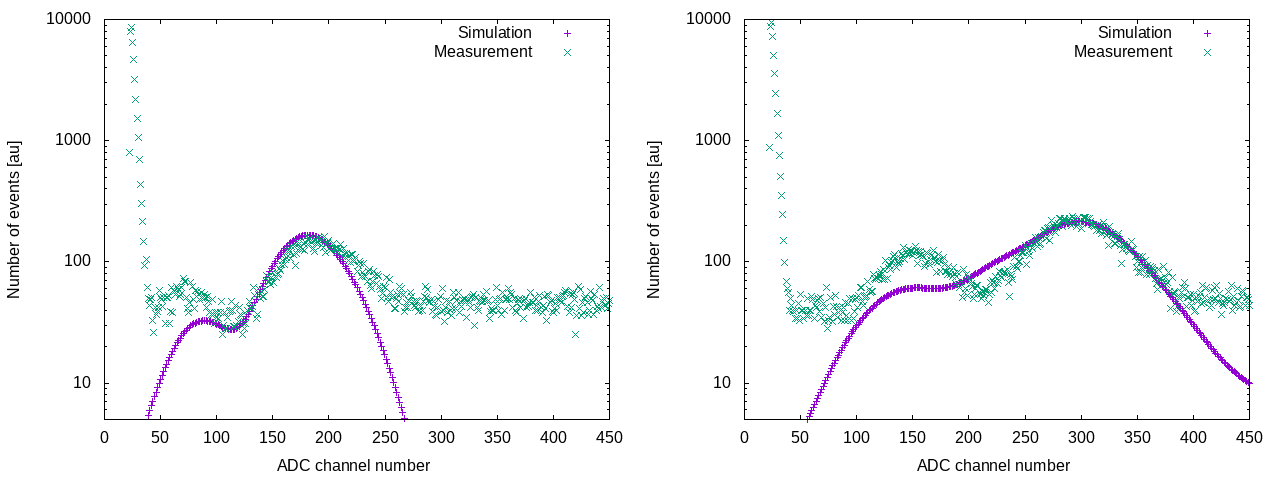}
 	}
 	\caption{Simulated spectra of number of scintillation photons detected compared to the measurements. At the closest position to the MPPC (right) and at one of the farthest corners (left). The simulated spectra were smeard by 5 and 15 channels respectively. \label{fig:spectra_meas}}
 \end{figure*}

 The measured and simulated spectra for the irradiation point closest to the MPPC and in the farthest corner are compared in Figure~\ref{fig:spectra_meas}. The same number of X-rays were simulated, which were emitted in 2 minutes of data acquisition for each spectrum. The difference between these spectra is the largest of all. The main reason for this is the difference in the mean path of optical photons, which is the shortest when the source is in front of the MPPC and the largest when the source is placed in the corner.

Two distinct peaks are visible in the measured and simulated spectra (Figure~\ref{fig:spectra_meas}). The peak with the higher energy corresponds to the $^{241}$Am $\gamma$ peak at 59.5\,keV. The lower-energy one is the K$_\alpha$ X-ray fluoresence peak of the Cesium in the scintillator \cite{Torigoe2019}. The results of the Geant4 simulation were smeared by a Gaussian function with a standard deviation $\sigma$ of 5 channels for the closest point of irradiation and 15 channels for the farthest to match them with the measured ones. The histogram of the number of photons detected in the simulation were scaled up by 1.35 and 1.39 respectively to match them with the measurements. This way assuming a linear detector response, all amplification factors were treated together. The fact that the scaling factor is almost the same for all parts of the scintillator translates to a good light collection efficiency. The main aim of the measurements was to determine the number of detected optical photons for an energy deposition of 1\,keV in the scintillator (on average). For the measurements taken at the farthest position from the MPPC the 59.5\,keV $^{241}$Am peak was at ADC channel 180. The number of detected photons in the simulations had to be scaled up by roughly 1.37 to match the measurements. For the measurements taken at the closest point to the MPPC the same scaling parameter was used. For the farthest position from the MPPC this implicates that an energy deposition of 1\,keV yields 4.11 detected optical photons on average.
 
 The light yield in the simulation for the scintillator was fixed at 54\,photons/keV, which is the yield for CsI(Tl) scintillators produced by Saint-Gobain\footnote{\url{https://www.crystals.saint-gobain.com/products/crystal-scintillation}} which is similar to our scintillator produced by AMCRYS\footnote{\url{http://www.amcrys.com}}. The absorption length and the reflectivity of the surface of the scintillator was varied until the simulation agreed with the measurements for the two extreme spectra, the one in front of the MPPC and the one in the farthest corner. The best fitting reflectivity was 99.99\,\% and the absorption length determined from the fit was 60\,cm. These values were used in the later simulations.

 In this way of calibration the energy resolution and noise of the electronics are taken into account in the simulations. Although pileup is not included but during operation we do not expect such high count rates from regular sources where it could be relevant.
 
\section{Description of Geant4 Simulations}
\label{sec:simulations}

A Geant4 MC based simulation was developed in order to understand how the \textit{CAMELOT} CubeSat constellation would detect $\gamma$-rays originating from short GRBs, long GRBs and TGFs. This required dedicated simulations of each background component as well as response to the $\gamma$-ray sources and calculation of the signal-to-noise ratio. The repository containing the simulation source code and analysis code are shared on GitHub (with a GNU General Public License). \footnote{\url{https://github.com/ggalgoczi/szimulacio/tree/master/Bck_4.10.6}}$^,$\footnote{ \url{https://github.com/ggalgoczi/szimulacio/tree/master/GRB}}.

As the first step, the experimental setup that was used to calibrate the optical parameters of the CsI(Tl) scintillator -- the $\gamma$-ray detector of the satellite -- that with its casing was implemented in Geant4. Details are in Sec.~\ref{sec:validate}. Afterwards the complex CAD model (Sec.~\ref{subseb:cad}) of the satellite -- consisting of 7 modules, each with a given average material composition -- was imported to Geant4 with CADMESH \cite{cadmesh}. Four scintillators, each read out by 8 MPPCs were placed on two sides of the satellite.

In order to keep computation time at a reasonable level, the simulation of each primary particle is stopped if the number of detected optical photons reaches 10\,000. Heavy ions can create several hundred thousands of scintillation photons. This limitation made it possible to run the codes on personal computers with a few cores. The signal of the 8 MPPCs is planned to be grouped into two groups of 4 MPPCs. In the following simulations the signal of all 32 MPPCs belonging to the four scintillators is summed up. This way we give a conservative signal-to-noise ratio (SNR) estimation since a more sophisticated trigger algorithm will decrease the chance of the background to exceed the threshold in each channel. The energy deposition is calculated from the number of detected optical photons. As described in Sec.~\ref{sec:validate}, 1\,keV energy deposition corresponds to 4.11 optical photons detected in the simulation. This corresponds to an overall photon detection efficiency of 7\% which is expected given the large size of the scintillator, the small sensitive area of MPPCs and the quantum efficiency of the MPPCs which depends on the scintillation light wavelength.

\subsection{Directional and Positional Distribution of Primary Particles}
\label{subsec:directional}

The simulations presented in this paper can be split into two main groups based on whether we are simulating the source of the background or an astrophysical source. The latter are the target objects: sGRBs, lGRBs and TGFs, which can be considered as point sources very far away, therefore photons coming from these sources are treated in the simulations as parallel.

The other main group is the background (eg. CXB, albedo particles, trapped electrons). The background particles and $\gamma$-rays mainly hit the satellite from large solid angles or isotropically. The pointing strategy of the \textit{CAMELOT} satellites is not established yet and the detector has all-sky FOV. We are interested in the estimation of a long-term average background at the regions of low geomagnetic latitude and outside SAA as mentioned in Sec.~\ref{sec:bck}. Therefore, as an approximation, we assume that all components of the background flux of particles and $\gamma$-rays irradiate the satellite isotropically.

In order to realize this in the simulations the background particles were placed randomly on a sphere with a radius $R$ around the model of the satellite. To maintain isotropy their direction was also randomly chosen. To boost up the simulation a source biasing was used \cite{gps} to limit the number of primary particles simulated to the ones which would hit the satellite. Due to the biasing, the number of detections in the simulation had to be normalized to determine the detection rate we would actually have.

The expected detection rate $N_\mathrm{det. rate}$ for \textit{CAMELOT} in space can be calculated as follows \cite{gps}:

\begin{equation}
N_\mathrm{det. rate} = 4 f_\Omega \pi^2 R^2 (\sin^2 \theta_\mathrm{max} - \sin^2 \theta_\mathrm{min}) \Phi N_\mathrm{det. sim}/N_\mathrm{prim} ,
\end{equation}

where $f_\Omega=\Omega/4\pi$ is the factor which takes into account the solid angle $\Omega$ of the type of the background. For instance albedo particles originate only from the atmosphere beneath the satellite. This corresponds to 3.95\,sr for our orbit. $R$ is the radius of the sphere upon which the primary particles are distributed. This radius needs to be much larger than the size of the target object to maintain isotropy. $\theta$ stands for the angle that is formed by the initial direction of the simulated primary particle and the vector pointing to the center of the satellite from the origin of the simulated primary particle. By limiting $\theta$ we are able to simulate only those particles which would hit the satellite. In our case $R=50$\,m. $\theta_\mathrm{min}$ and $\theta_\mathrm{max}$ are the upper and lower bounds for the chosen interval of the emission angle in the simulation. In our case $\theta_\mathrm{min}=0$ and $\theta_\mathrm{max}=0.5729 ^{\circ}$. $\Phi$ is the flux in units of cm$^{-2}$s$^{-1}$sr$^{-1}$. $N_{\mathrm{det. sim}}$ is the number of detections in the given simulation. $N_\mathrm{prim}$ is the number of primary particles shot in the simulation. The factor $\pi^2 R^2 (\sin^2 \theta_\mathrm{max} - \sin^2 \theta_\mathrm{min}) = 24668$\,cm$^2$. Table~\ref{tab:norm} summarizes the values of $\Omega$ and the normalization factor $f_\mathrm{norm} = 4 \Omega \pi^2 R^2 (\sin^2 \theta_\mathrm{max} - \sin^2 \theta_\mathrm{min})$ for the background models described in Sec.~\ref{sec:bck}.

\begin{table}[h]
\caption{Summary of solid angles $\Omega$ of background flux and normalization factor $f_\mathrm{norm}$.}
\begin{center}
\begin{tabular}{|c|c|c|c|c|}
\hline
\rule[-1ex]{0pt}{3.5ex}                              & CXB and    & Galactic & Trapped and charged  & Albedo             \\
\rule[-1ex]{0pt}{3.5ex}                              & primary CR & $\gamma$ & secondary particles  & $\gamma$ and n$^0$ \\
\hline
\rule[-1ex]{0pt}{3.5ex} $\Omega$ (sr)                & 8.64       & 0.542    & $4\pi$               & 3.93                \\ 
\rule[-1ex]{0pt}{3.5ex} $f_\mathrm{norm}$ (cm$^2$sr) & $8.525\times 10^5$ & $5.348\times 10^4$ & $1.240\times 10^6$ & $3.878\times 10^5$\\
\hline
\end{tabular}
\end{center}
\label{tab:norm}
\end{table}

%\subsection{CAD model}
\subsection{Satellite's Mass Model}
\label{subseb:cad}

In order to include all parts of the satellite including even the smallest volumes, the detailed CAD model of the satellite was read into Geant4 directly with CADMESH \cite{cadmesh} that utilizes TETGEN \cite{tetgen} and ASSIMP\footnote{\url{http://www.assimp.org}} software libraries to directly read in STL files into Geant4. The satellite consists of seven modules including the structure of the satellite, the communications module, the payload etc. The only volume that was not included were the antennas. The material of each of the volumes was averaged (as described in Table~\ref{table:materials1}. and Table~\ref{table:materials2}.). The complete list of alloys used in each volume are listed in Sec.~\ref{appendix_a} together with their composition. Figure~\ref{fig:mass_model} shows the mass model of the \textit{CAMELOT} satellite. Figure~\ref{fig:themodel} in the Appendix presents the individual volumes of the satellite.

\begin{figure}[h]
\begin{center}
\begin{tabular}{c}
\includegraphics[width=0.8\linewidth]{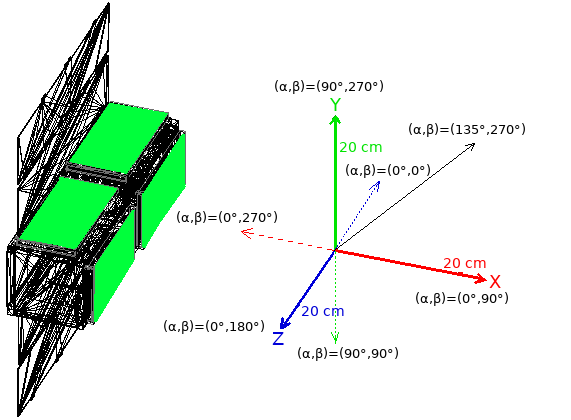}
\end{tabular}
\end{center}
\caption 
{ \label{fig:mass_model}
The mass model used for the Geant4 simulations. The CAD model of the satellite was read into Geant4 and 4 scintillators (green rectangles) with their respective read out were placed on two sides of the satellite. The angles $\alpha$ and $\beta$ refer to the angles shown in the figure of the detector's effective area. $\beta$ is rotation around the Y axis, counted from the -Z axis and it increases towards +X axis.
$\alpha$ is rotation around the Z axis, counted from the +X axis and increasing towards -Y axis.
The highest effective area is for a source at direction ($\alpha$, $\beta$) = (135$^\circ$, 270$^\circ$).}
\end{figure}

\section{Results of Geant4 Simulations}
\label{sec:sim_results}

\subsection{Response to Each External Background Components}
\label{sec:sim_background}

\begin{figure}[h]
\begin{center}
\begin{tabular}{c}
\includegraphics[width=0.9\linewidth]{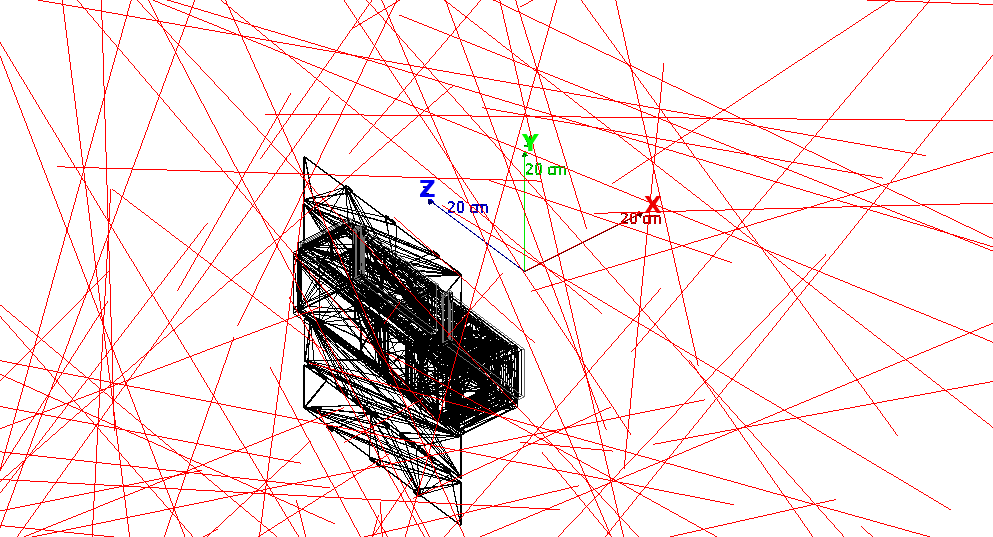}
\end{tabular}
\end{center}
\caption 
{ \label{fig:themodel_iso_rad}
The mass model of the satellite is isotropically irradiated with X-rays (blue tracks). All four scintillators onboard appear green as they are filled with the tracks of optical photons which have green colour. The effect of the directional biasing (described in Sec.~\ref{subsec:directional}) can be seen.} 
\end{figure}

In this section the simulation results of the satellite response to each of 14 external background components is presented. The results are for one \textit{CAMELOT} satellite. The count rate is summed for all detectors. By far the most relevant background is CXB. Therefore we chose to simulate two different CXB models introduced by Gruber et al. (1999) \cite{Gruber1999} and Ajello et al. (2008) \cite{Ajello2008} described in Sec.~\ref{sec:cxb}. The input energy spectra of each background component used for the simulations are described in the corresponding subsection of section \ref{sec:bck}. The model of the satellite was irradiated isotropically as shown in Figure~ \ref{fig:themodel_iso_rad}.
 
Four possible aluminium detector support structure (shortly detector casing or just casing) thicknesses were investigated. The same material and thickness is on all sides of the detector housing, including the back side. In order to give an idea of the contribution of each component a realistic 20\,keV low-energy threshold was chosen. The five components which contribute the most to the background for the casing thickness of 0.5\,mm thick Al with this low-energy threshold are: CXB ($\sim 1000-1100$ counts per second (cps)), albedo $\gamma$ ($\sim 200$\,cps), primary CR protons (27\,cps), albedo protons (45\,cps) and albedo positrons (28\,cps). Tables~\ref{table:count_rates1} and \ref{table:count_rates2} summarize the background detection rate predicted by the simulation.
 
\begin{table}[h]
\caption{Simulated detection rate induced by cosmic and trapped particle background components.}
\begin{center}
\begin{tabular}{|c|c|c|c|c|c|c|c|c|c|}
\hline
Thickness & CXB & CXB & CR & CR & Galactic & Trapped & CR  & CR & Trapped \\
(mm)      & A08 & G99 & $\alpha$ & p$^{+}$ & $\gamma$ & p$^{+}$ & e$^{-}$ & e$^{+}$ & e$^{-}$  \\
\hline
0.5 & 1150 & 996 & 51 & 28 & 5.12 & 1.15  & 0.74 & 0.057 & 0.17 \\
1.0 & 1020 & 893 & 49 & 29 & 3.98 & 0.947 & 0.76 & 0.057 & 0.073 \\
1.5 & 890  & 770 & 51 & 29 & 4.04 & 0.820 & 0.76 & 0.060 & 0.072 \\
2.0 & 858  & 707 & 51 & 29 & 3.50 & 0.827 & 0.75 & 0.059 & 0.066 \\
\hline
\end{tabular}
\end{center}
The background detection rate is in counts per second, assumes a low-energy threshold of 20\,keV, and is simulated for different thicknesses of the aluminium support structure of the detector. Two spectral models described by Gruber et al. (1999) \cite{Gruber1999} and by Ajello et al. (2008) \cite{Ajello2008} (denoted as G99 and A08) were simulated for the CXB. For the primary CR p$^{+}$ and $\alpha$ particles we used the ISO-15390 model with stormy magnetosphere and inclination of $i=20^\circ$. For primary CR e$^{-}$ and e$^{+}$ we used the model described by Mizuno et al. (2004) \cite{Mizuno2004} for solar minimum, $\theta_\mathrm{M}=29.6^\circ$. For trapped e$^{-}$ and p$^{+}$ we used the AE9 and AP9 models, respectively, for inclination of $i=20^\circ$, MC mode and derived from the 50\,\% CL of the fluxes. Altitude of 500\,km was chosen.
\label{table:count_rates1}
\end{table}

\begin{table}[h]
\caption{Simulated detection rate induced by background components originating in the atmosphere.}
\begin{center}
\begin{tabular}{|c|c|c|c|c|c|}
\hline
Thickness (mm) & Albedo $\gamma$ &	Secondary e$^{+}$ &	Secondary e$^{-}$ &	Secondary p$^{+}$ &	Albedo n$^{0}$  \\
\hline
0.5 & 208 & 28.2 & 6.32 & 45.5 & 23.8  \\
1.0 & 205 & 27.1 & 8.00 & 43.7 & 22.3  \\
1.5 & 192 & 27.4 & 7.95 & 42.5 & 22.3  \\
2.0 & 191 & 28.2 & 7.86 & 43.4 & 21.4  \\
\hline
\end{tabular}
\end{center}
The detection rate is in counts per second, assumes a low-energy threshold of 20\,keV and is simulated for different thicknesses of the aluminium support structure of the detector. For secondary neutrons the spectrum for the solar minimum and for the cutoff rigidity $R_\mathrm{cut}=5$\,GV was used.
\label{table:count_rates2}
\end{table}

By summing up the contribution of each background component we derived a total background rate of 1550\,cps for 0.5\,mm, 1400\,cps for 1\,mm, 1270\,cps for 1.5\,mm and 1100\,cps for 2\,mm of the casing thickness assuming a low-energy threshold of 20\,keV. From the two CXB models simulated, the Ajello et al. (2008) model was chosen since it gives a more conservative estimate. The Ajello et al. (2008) and Gruber et al. (1999) models have integral fluxes of 33.7\,ph\,cm$^{-2}$s$^{-1}$ and 30.3\,ph\,cm$^{-2}$s$^{-1}$ for $E>10$\,keV and give background rates of 1020\,cps and 893\,cps for 1\,mm thick Al detector casing, respectively.

\begin{figure}[h]
\begin{center}
\begin{tabular}{c}
\includegraphics[width=0.7\linewidth]{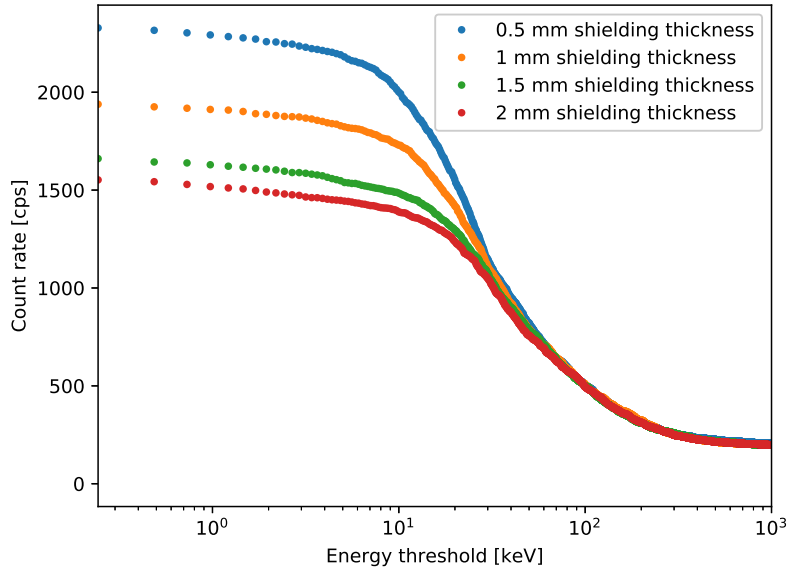}
\end{tabular}
\end{center}
\caption 
{Background count rate for three aluminium detector support structure thicknesses versus low-energy thresholds. Above a threshold of $\sim$50\,keV the thickness of the support structure does not change the background rate. \label{fig:threshold_vs_back}} 
\end{figure}

The low-energy threshold onboard the \textit{CAMELOT} satellites is planned to be a tunable parameter and changeable upon a ground command. Laboratory experiments show that low-energy threshold for our detectors is around $15-20$\,keV. In Figure~\ref{fig:threshold_vs_back} the detection rate is shown for different casing thicknesses and low energy thresholds. The low energy part of the background spectrum is dominated by CXB X-rays which are stopped by thicker detector casing. The higher energy part is dominated mostly by hadrons and albedo gamma rays (hard spectrum) which can easily cross aluminium and deposit high energies in the scintillator.

\subsection{Simulation of a Typical Short GRB from Different Directions}
 
In order to quantify if \textit{CAMELOT} is capable of detecting X-ray sources we need to investigate the X-ray absorption by the satellite structures themselves. The four scintillators are placed on two sides of the satellite. Therefore the X-rays from half of the objects need to cross a certain part of the satellite before arriving to the detectors.
 
 \begin{figure}[h]
\begin{center}
\begin{tabular}{c}
\includegraphics[width=0.9\linewidth]{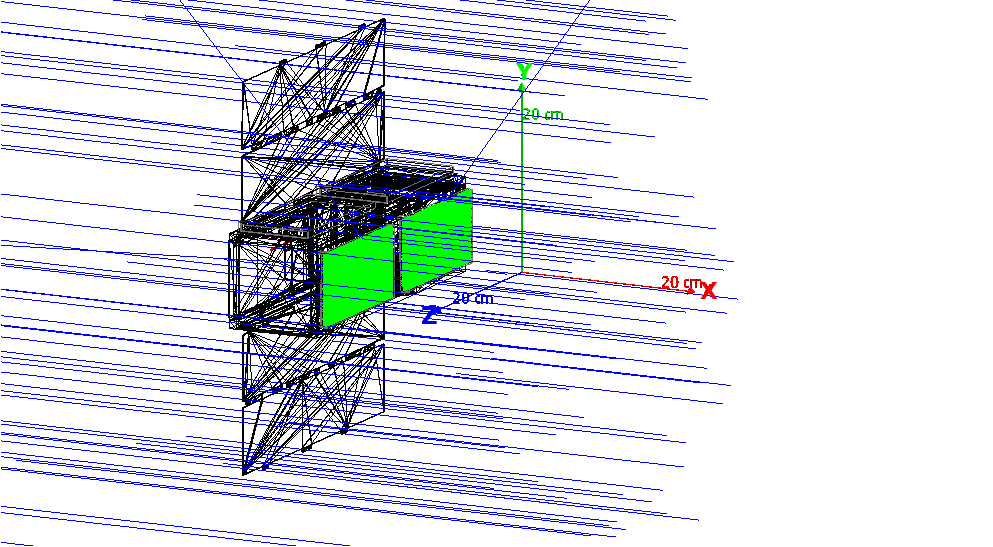}
\end{tabular}
\end{center}
\caption 
{ \label{fig:theparallel}
The satellite hit by parallel X-rays, such as the ones originating in short GRBs. Only one of the four scintillators were triggered in this case. Blue tracks are X-rays, green tracks are optical photons which were detected by the MPPCs.} 
\end{figure} 
 
To quantify the X-ray absorption of the satellite and to determine the count rate expected from sGRBs, different source directions were simulated. First, the satellite was rotated around its major axis by 10$^{\circ}$ 35 times to cover all directions around this axis. Afterwards the same was repeated for the minor axis of the satellite. A 1024\,ms peak spectrum of a typical sGRB was used in the simulation. Details of typical sGRB spectra are described in subsection \ref{sec:GRB}.

\subsubsection{Rotating Around Major Axis for Short GRB}

\label{subsec:major_axis}
\begin{figure}[h]
 \begin{center}
 \begin{tabular}{c}
 \includegraphics[width=0.7\linewidth]{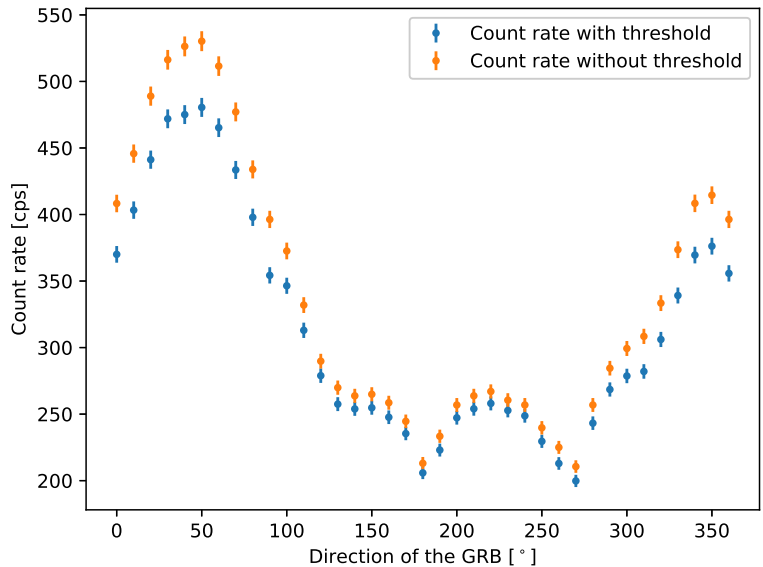}
 \end{tabular}
 \end{center}
 \caption 
 {Count rate of a typical sGRB (for 1024\,ms peak spectrum) for different source directions. The satellite was rotated around its major axis. An arbitrary but possible low-energy threshold of 20\,keV was utilized. The detector support structure thickness of the scintillator was 2\,mm. \label{fig:diff_direction_major_axis}
  } 
 \end{figure}

The satellite was rotated around its major axis (Z in Figure~\ref{fig:theparallel}) by $10^{\circ}$ between each simulation. The simulated primary X-rays originated from the direction of the X axis. $0^{\circ}$ case corresponds to the scenario when X-rays arrive perpendicular to the surface of two of the scintillators. The count rate is the highest for $45^{\circ}$ when the angle between the direction of the photons and the surface normal of both detectors is $45^{\circ}$. The combined projected area of the four scintillators is the largest in this scenario. The least favored direction is $270^{\circ}$ (see Figure~\ref{fig:diff_direction_major_axis}).

\begin{figure}[h]
 \begin{center}
 \begin{tabular}{c}
 \includegraphics[width=0.7\linewidth]{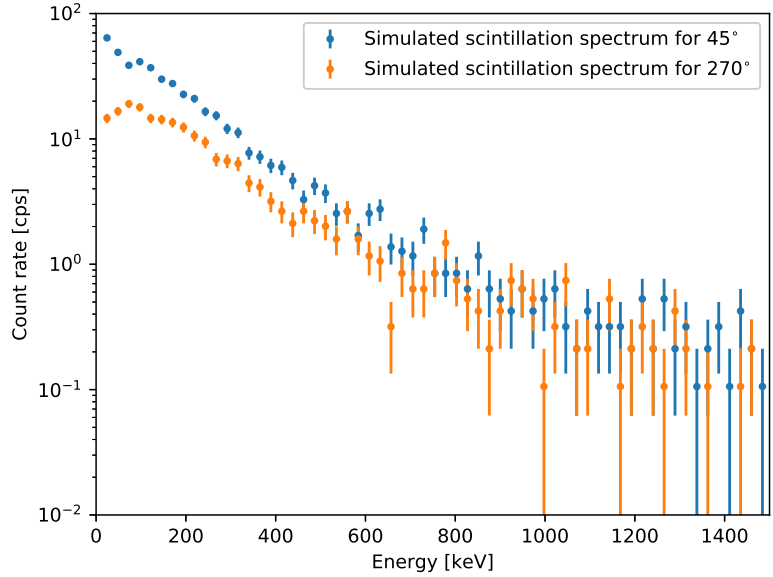}
 \end{tabular}
 \end{center}
 \caption 
 {\label{fig:diff_direction_spectra}
 Typical short GRB spectra (for 1024\,ms peak) for the most optimal direction and the least optimal direction among the investigated cases. The lower energy band is suppressed for 270$^{\circ}$, since GRB X-rays need to cross the material of the satellite for this scenario to be detected. Thickness of the detector support structure was 2\,mm in this case.} 
 \end{figure}

In Figure~\ref{fig:diff_direction_spectra} the spectra of two directions are shown. As expected for the least optimal direction ($270^{\circ}$) the low energy part of the spectrum is suppressed. These are the X-ray photons which are not able to cross the material of the satellite.

\begin{figure}[h]
 \begin{center}
 \begin{tabular}{c}
 \includegraphics[width=0.7\linewidth]{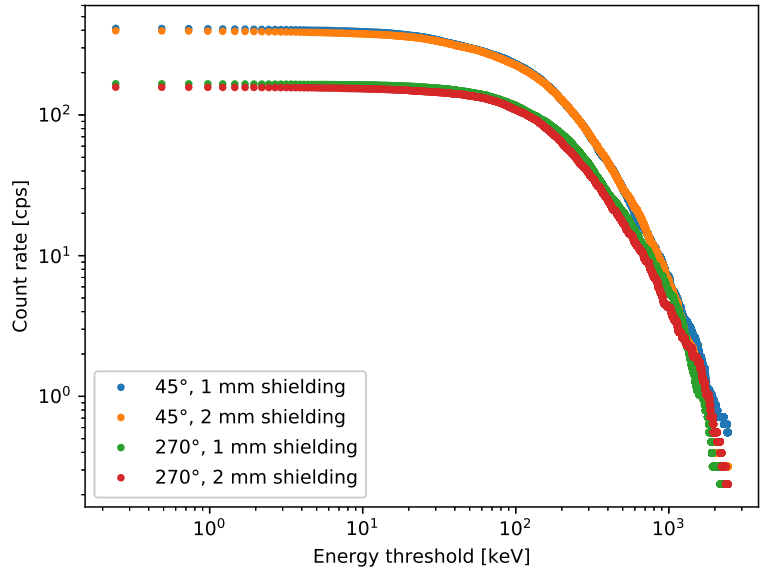}
 \end{tabular}
 \end{center}
 \caption 
 {\label{fig:det_vs_threshold}
 Count rate for a typical sGRB (for 1024\,ms peak spectrum) versus detection low-energy threshold. Two source directions and thicknesses of the detector support structure are shown from the 70 investigated scenarios.}
 \end{figure}

 Different low-energy thresholds are possible to be set. Therefore, it is important to understand how the count rate of X-rays from sGRBs would change by varying the low-energy threshold. In Figure~\ref{fig:det_vs_threshold} the count rate for the most and least optimal direction is shown depending on the low-energy threshold. Up to about 100\,keV the count rate does not decrease significantly. Also it is important to notice that the direction of the source is much more important than the thickness of the detector casing.
 
 \subsubsection{Rotating Around Minor Axis for Short GRBs}

\begin{figure}[h]
 \begin{center}
 \begin{tabular}{c}
 \includegraphics[width=0.7\linewidth]{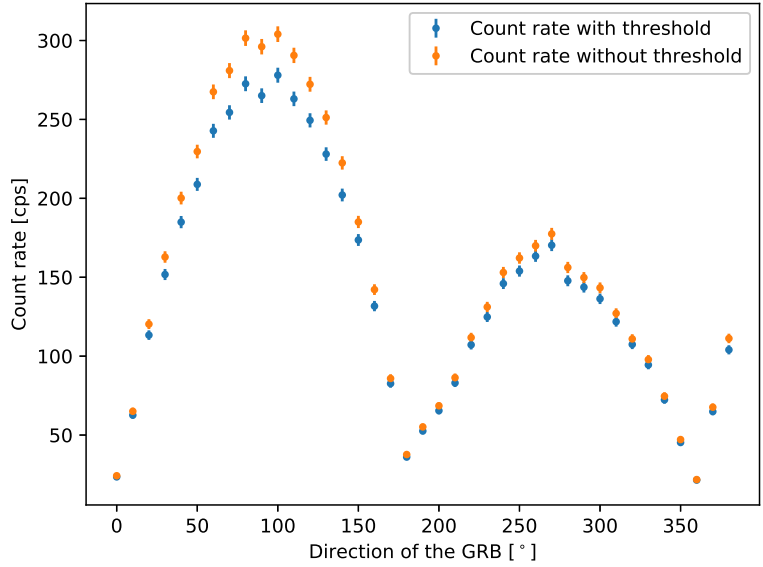} 
 \end{tabular}
 \end{center}
 \caption 
 {\label{fig:det_vs_direction_minor_axis}
 Count rate of a typical sGRB (for 1024\,ms peak spectrum) for different source directions. The satellite was rotated around its minor axis. An arbitrary but possible low-energy threshold of 20\,keV was utilized. The aluminium detector support structure was 2\,mm thick in this case.}
 \end{figure}

The same procedure as described in subsection \ref{subsec:major_axis} was followed for investigating sGRB (for 1024\,ms peak spectrum) directions around the minor axis (X in Figure~\ref{fig:theparallel}). X-ray were simulated as a parallel beam coming from the Z direction. In Figure~\ref{fig:det_vs_direction_minor_axis} the count rate for each direction is shown. $0^{\circ}$ is the least optimal. It corresponds to the case when all scintillators are seen from their edge. In this case the cross-section of the four scintillators combined is 7.5\,cm$^{2}$, which is about 3\,\% of the area in the most optimal direction during rotation around the minor axis.

 Since the background for the casing thickness of 2\,mm is 1100\,cps, a count rate of 165\,cps is required from a sGRB to be observed with a significance of 5\,$\sigma$. Therefore from the directions investigated with the rotation of the minor axis the interval between $50^{\circ}$ and $150^{\circ}$ is suitable for the detection of sGRBs.
 
\subsection{Signal-to-Noise Ratio of X-ray/$\gamma$-ray Transients}
 
 For the detection of astrophysical objects, the final figure of merit is the SNR. It is important to mention that for the localization accuracy not only the SNR but also the number of detected photons is important for the cross-correlation of light curves. Electronic noise was neglected in the following calculations as the planned low-energy threshold set for detection is higher than the amplitude of the electronic noise.
 
Detection low-energy threshold can be set onboard the satellite. Therefore SNR was quantified for each astrophysical object in the function of low-energy threshold. The following equation was used to determine SNR:

\begin{equation}
 SNR = \frac
{ \Delta t \sum\limits_{E_1}^{E_2} f(E) }
{ \sqrt{\Delta t \sum\limits_{E_1}^{E_2} g(E) } },
\end{equation}

where $f(E)$ is the detected count rate spectrum of the signal, $g(E)$ stands for the detected count rate background spectrum, $E_\mathrm{1}$ is the low-energy threshold, $E_\mathrm{2}$ is the high-energy threshold and $\Delta t$ is the exposure time. In our simulations we did not put boundary on the high-energy threshold, but the triggering algorithm onboard \textit{CAMELOT} satellites will have capability to set both the low- and the high-energy thresholds.

The main aim of the \textit{CAMELOT} mission is the detection and localization of sGRBs. Therefore it is important to understand with what significance could \textit{CAMELOT} satellites detect these objects. Among the direction, when the satellite was rotated around its major axis SNR is the highest when the angle between X-rays from a GRB and the surface normal of both detectors is $45^{\circ}$. We have the lowest SNR for the angle of $270^{\circ}$ (see Figure~\ref{fig:diff_direction_major_axis}). 
 
 \begin{figure}[h]
 \begin{center}
 \begin{tabular}{c}
 \includegraphics[width=0.7\linewidth]{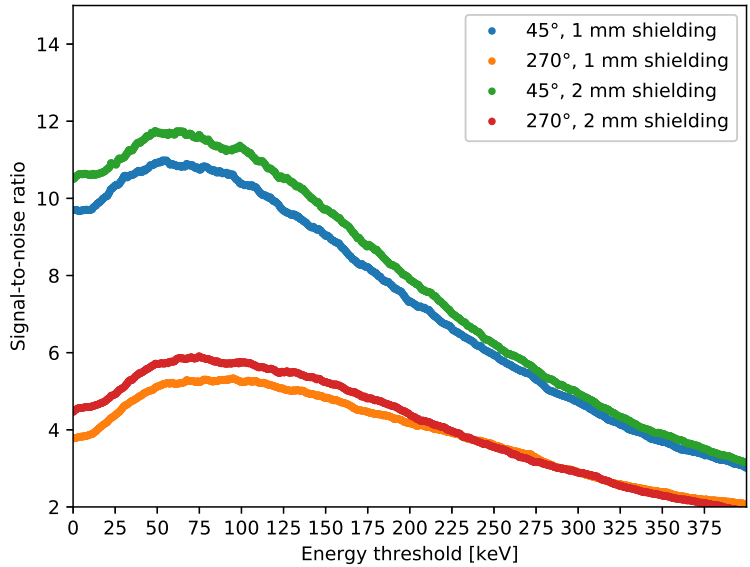}
 \end{tabular}
 \end{center}
 \caption 
 { \label{fig:snr_energy_htreshold}
 Signal-to-noise ratio of short GRBs versus low-energy detection threshold. Two detector support structure thicknesses and two GRB directions (rotating around Z axis) are shown among the directions investigated in Sec.~$\ref{subsec:major_axis}$. Among these investigated directions, highest signal is achieved when the direction of the GRB is $45^{\circ}$. For $270^{\circ}$ the signal from a typical sGRB is the smallest (see Figure~\ref{fig:diff_direction_major_axis}).}
 \end{figure} 
 
 When the simulated typical sGRB (for 1024\,ms peak spectrum) is in the most optimal direction an SNR of $>11$ can be achieved. SNR stays above 10 within the low-energy detection threshold range from up to 100\,keV. The thickness of the aluminium detector casing affects SNR mostly for low values of the low-energy detection thresholds, since the main background component, CXB has a rather soft spectrum. For the least optimal direction among the directions investigated an SNR of 6 can be achieved. This characteristic is shown in Figure~\ref{fig:snr_energy_htreshold}. The other directions, when the satellite was rotated around its minor axis (Figure. \ref{fig:det_vs_direction_minor_axis}), are less favored.
 In these cases the cross section of the detectors are significantly lower.
 
 Tables~\ref{table:count_rates_transients} and \ref{table:snr} summarize the simulated detection count rate induced by the X-ray/$\gamma$-ray transient sources and the expected SNR. For GRBs four different exposure times $\Delta t$ were used: 64, 256 and 1024\,ms for sGRBs and 4096\,ms for lGRBs. It should be noted that long triggering timescales of the order of several seconds or tens of seconds are readily affected by the time variation of background due to geomagnetic latitude (cutoff rigidity) change during the orbit and the activation background varying with time since SAA passages. The SNR calculated for such long integration time is effected by the background systematics. Varying background can cause false triggers and detectors with larger effective area are more vulnerable unless a sophisticated background modeling is part of the trigger algorithm.
 
 Different missions has employed different triggering timescales. \textit{BeppoSAX}/GRBM used adjustable timescale in the range from 7.8125\,ms to 4\,s in 10 steps \cite{Feroci1997}. \textit{CGRO}/BATSE used time windows of 64 ms, 256\,ms and 1024\,ms \cite{Band2002}. \textit{Suzaku}/WAM used triggering timescales of 1/4 \,s and 1\,s \cite{Yamaoka2017}. \textit{HETE-2}/WXM and FREGATE used timescales from 80\,ms up to 10.5 s or longer, but they modeled background to remove trends which can cause false triggers \cite{Fenimore2001}. \textit{Swift}/BAT uses two types of rate triggers: i) ``short" rate triggers with timescales 4, 8, 16, 32 and 64\,ms which are traditional triggers employing single background period of fixed duration; ii) ``long" rate triggers with timescales from 64\,ms to 64\,s which fit multiple background intervals to remove trends as pioneered by \textit{HETE-II} \cite{Fenimore2003}. \textit{Fermi}/GBM uses triggering timescales of 16, 32, 64, 128, 256, 512, 1024, 2048, 4096 and 8192\,ms \cite{NarayanaBhat2016}. In case of \textit{AGILE}/MCAL, transients are searched using time windows of duration of sub-millisecond, 1, 16, 64 , 256, 1024 and 8192\,ms \cite{Feroci2007, Fuschino2008}.
 
 From Table~\ref{table:snr} it is also seen that in contrary to one's expectation, TGF SNR increases with thicker detector casing. The reason for this is the hardness of TGF spectrum. The median energy of X-ray photons originating in TGFs is 2\,MeV. These have a few per cent chance to interact with the scintillator material. The thicker detector casing provides more material in which these energetic photons can interact and produce secondary particles. Therefore thicker casing yields in higher SNR. This study is important in order to understand sensitivity of different trigger time window durations necessary for designing efficient GRB trigger algorithm.

\begin{table}[h!]
\caption{Simulated detection rate induced by X-ray/$\gamma$-ray transient sources.}
\begin{center}
\begin{tabular}{|c|c|c|c|c|c|c|c|c|c|}
\hline
Al   & \multicolumn{3}{c|}{sGRB peak spectrum} & \multicolumn{3}{c|}{lGRB peak spectrum} & lGRB & \multirow{2}{*}{TGF} & \multirow{2}{*}{SGR} \\ \cline{2-7}(mm) & 64\,ms & 256\,ms & 1024\,ms &    64\,ms & 256\,ms & 1024\,ms & fln. sp. &      &  \\\hline
0.5 & 1440 & 910 & 385 & 1190 & 924 & 758 & 326 & 34600 & 16200 \\ 
1.0 & 1390 & 911 & 367 & 1139 & 908 & 715 & 309 & 38900 & 14000 \\ 
1.5 & 1300 & 890 & 367 & 1110 & 840 & 674 & 292 & 37600 & 11900 \\ 
2.0 & 1320 & 839 & 355 & 1058 & 815 & 662 & 277 & 38600 & 10500 \\ 
\hline
\end{tabular}
\end{center}
The detection rate is in counts per second, assumes a low-energy threshold of 20\,keV and is simulated for different thicknesses of the aluminium support structure of the detector. For short and long GRBs the 64\,ms, 256\, ms and 1024\,ms peak spectra were used. For long GRB also the fluence spectrum (fln. sp.) was used.
\label{table:count_rates_transients}
\end{table}

\begin{table}[h]
\caption{Simulated detection signal-to-noise ratio for X-ray/$\gamma$-ray transient sources.}
\begin{center}
\begin{tabular}{|c|c|c|c|c|c|c|c|c|c|}
\hline
Al   & \multicolumn{3}{c|}{sGRB peak spectrum} & \multicolumn{3}{c|}{lGRB peak spectrum} & lGRB fln. sp. & TGF & SGR \\
\cline{2-10}
(mm) & 64\,ms & 256\,ms & 1024\,ms &  64\,ms & 256\,ms & 1024\,ms & 4096\,ms & 0.1\,ms & 0.2\,s \\\hline
0.5  & 9.27 & 11.7 & 9.91 & 7.66 & 11.9 & 19.5 & 16.8 & 8.79 & 185 \\ 
1.0  & 9.35 & 12.3 & 9.87 & 7.66 & 12.2 & 19.2 & 16.7 & 10.3 & 167 \\ 
1.5  & 9.23 & 12.6 & 10.4 & 7.89 & 11.9 & 19.2 & 16.6 & 10.6 & 149 \\ 
2.0  & 10.1 & 12.8 & 10.8 & 8.08 & 12.5 & 20.2 & 16.0 & 11.6 & 142 \\ 
\hline
\end{tabular}
\end{center}
The detection SNR has been calculated for different thicknesses of the aluminium support structure of the detector. The assumed background count rate is the sum of all components as described in Ref.~\ref{sec:sim_background}. For the GRB peak spectra were used the exposure time $\Delta t=$ 64\,ms, 256\,ms and 1024\,ms. For the fluence spectrum (fln. sp.) of long GRB we used $\Delta t=4096$\,ms. For TGF we assume exposure time $\Delta t=0.1$\,ms and for SGR $\Delta t=0.2$\,s. For this TGF spectrum the SNR is only a theoretical value following from the formula. For example, for 1\,mm thicknesses of the Al detector casing the expected detected number of counts within 0.1\,ms from is 3.9\,cnt, whereas the expected number of background counts is below 1\,cnt (only 0.14\,cnt).
\label{table:snr}
\end{table}

\section{Discussion}
\label{sec:discuss}

The background count rates obtained in our MC simulations were derived from models of fluxes of gamma-rays and particles averaged over various latitudes, depending on the particular flux component, below $50^\circ$ and outside SAA. The reason is that we aim to obtain an expected ``mean" background rate in parts of orbit which are suitable for gamma-ray transient scientific data collection. However, in reality, the background rate will have time variation due to geomagnetic latitude (cutoff rigidity) change within the orbit and due to the activation background varying with time since SAA passages.

The foreseen orbital inclination of the \textit{CAMELOT} satellites is above $\sim 50^\circ$ and option of polar orbits is also likely. Therefore significant background variation is expected as well. We investigated background count rates measured by \textit{Fermi}/GBM, \textit{RHESSI}\cite{Smith2002} and \textit{Lomonosov}/BDRG\cite{Svertilov2018} gamma-ray instruments throughout their orbits to learn what background variations are expected. In case of \textit{Fermi}/GBM (altitude 560\,km and inclination 26$^{\circ}$) the background rate outside SAA for one NaI detector module in $8-1000$\,keV range varied throughout the orbit between $\sim 1000$\,cps and $\sim 1500-2000$\,cps, i.e. $\sim 1.5-2\times$ change (about two months after the launch), see also Ref.~\citenum{Fitzpatrick2012}. In case of the \textit{RHESSI} spectrometer (altitude 500\,km and inclination 38$^{\circ}$) the background rate outside SAA for one rear detector segment in $25-20\,000$\,keV range varied throughout the orbit between $\sim 250$\,cps and $\sim 650$\,cps, .i.e. $\sim 2.5\times$ change (about three years after the launch) \footnote{\url{http://sprg.ssl.berkeley.edu/~tohban/browser/}}. For a detector on a polar orbit the background rate can increase dramatically more when passing the polar regions of trapped particles in the van Allen radiation belt. The \textit{Lomonosov}/BDRG (altitude 550\,km and inclination 98$^{\circ}$) measurements show that the rate outside SAA, in $20-450$\,keV range increases $\sim 50\times$ inside the polar regions compared to the rates near equator (about 5 months after the launch), see also Ref.~\citenum{Svertilov2018}. Therefore outside SAA and polar trapped particle regions the background count rate variation of at least $\sim 2\times$ the value near the equator for detectors on board the \textit{CAMELOT} satellites are also foreseen. Inside the polar regions the rate increase can be much higher.

Since CubeSats which are not on equatorial orbit (currently most of them) are subject to high proton flux upon SAA passages, these protons have enough energy to activate the material of the satellite \cite{Cumani2019}. Short term activation is important as decaying isotopes can cause a strongly time-variable background for a few minutes after the end of SAA passes. Long term activation can increase the background significantly in a matter of months. \cite{Kokubun1999} To quantify the effects of proton induced activation, we plan to conduct simulations in the near future. Emission lines of radioactive isotopes could also be used for energy calibration. As an example for our scintillators $^{123}$I will be created which emit $\gamma$-rays with an energy of 159\,keV \cite{Gruber1989}.

In order to discuss how our estimated background rates of \textit{CAMELOT} detectors scale to the observations of \textit{Fermi}/GBM, \textit{AGILE}/MCAL and \textit{Suzaku}/WAM we consider the surface area of the scintillators of these instruments and assume that these surface areas can be used as rough scaling factors.
For CAMELOT (CsI): $15\,\mathrm{cm} \times 7.5\,\mathrm{cm} \times 4$ scintillators giving the area of $\sim 450$\,cm$^2$. For \textit{Fermi}/GBM (NaI): $3.14 \times (12.7/2\,\mathrm{cm})^2$ giving the area of $\sim 127$\,cm$^2$ for one detector module (the effective area around 0.4\,MeV is $\sim 30$\,cm$^2$) \cite{Meegan2009}. For \textit{AGILE}/MCAL (CsI): $1.5\,\mathrm{cm} \times 37.5\,\mathrm{cm} \times 30$ detectors giving the area of $\sim 1700$\,cm$^2$ (the effective area at $0.4-1$\,MeV is $\sim 200-300$\,cm$^2$) \cite{Labanti2009}. For \textit{Suzaku}/WAM (BGO) the area is $\sim 800$\,cm$^2$ \cite{Yamaoka2017}.

The estimated background rates of \textit{CAMELOT} detectors for 0.5\,mm Al support structure are $\sim 2.3$, $\sim 0.75$ and $\sim 0.3$\,kHz respectively for $E>10$\,keV, $E>50$\,keV and $E>400$\,keV (see Fig.~\ref{fig:threshold_vs_back}). The observed background rates of one detector module of \textit{Fermi}/GBM ($\sim 2$ months after the launch, altitude $h=560$\,km and inclination $i=26^\circ$) are $\sim 1$\,kHz and $\sim 0.13$\,kHz for $E>10$\,keV and $E>400$\,keV, respectively. For \textit{AGILE}/MCAL ($h=550$\,km and $i=2.5^\circ$) the rate is $\sim 1.3$\,kHz for $E>400$\,keV. In case of \textit{Suzaku}/WAM ($h=570$\,km and $i=31^\circ$) the rate is $4-6$\,kHz for $E>50$\,keV.

If we use the surface area of scintillators as a scaling factor than the background rate scaling between \textit{Fermi}/GBM and \textit{Suzaku}/WAM or \textit{Fermi}/GBM and \textit{AGILE}/MCAL is in a good approximation. The current simulated \textit{CAMELOT} background is smaller than that of \textit{Fermi}/GBM and \textit{Suzaku}/WAM observations. For example, if we scale the \textit{CAMELOT} background to the \textit{Fermi}/GBM, it would be $\sim 0.65$\,kHz, which is lower than the observed value of $\sim 1$\,kHz for $E>10$\,keV. This is of course due to the activation background component, which is not included in the \textit{CAMELOT} simulated background and the scaling would be reasonable if we consider the activation background component for \textit{CAMELOT} around $1-2$\,kHz. The same applies when we scale \textit{CAMELOT} background to \textit{Suzaku}/WAM observations. Scaling \textit{CAMELOT} background to \textit{AGILE}/MCAL is in a good agreement even without accounting for the activation component for \textit{CAMELOT}. The \textit{AGILE}/MCAL satellite is in an equatorial orbit with an inclination of only $2.5^\circ$ with low particle background which causes material activation.

Having the simulation results of the detection background count rate and GRB count rate we can discuss an approximate number of expected short and long GRB detections per year by a single \textit{CAMELOT} satellite. For 1\,mm Al casing and the best gamma-ray incident angle ($45^{\circ}$ to the surface normals of perpendicular detectors) the simulation detection count rate is 911\,cps and 715\,cps for sGRB 256\,ms and lGRB 1024\,ms median peak spectrum, respectively. GRBs will not always be seen under this most preferred direction therefore we scale this rate by a factor of $1/\sqrt{2}$ which corresponds to the rate expected from GRBs seen perpendicular to the largest scintillator side (644\,cps for sGRB and and 506\,cps for lGRB). There will be GRBs seen under more preferred direction as well as under less preferred direction therefore this is a compromise direction. Furthermore we can assume a detection SNR threshold to be 5 and a mean background count rate to be 3\,000\,cps (1\,500\,cps from external background flux and 1\,500\,cps from material activation).

Therefore for the above-mentioned integration times we obtain detection count rate thresholds of 541\,cps and 271\,cps, respectively for sGRBs and lGRBs with median spectral shapes. By scaling the spectral normalizations $A_{256}$ and $A_{1024}$ of typical \textit{Fermi}/GBM sGRB and lGRB from Table~\ref{tab:grb_param} by factors of
$541/644$ for sGRB and by $271/506$ for lGRB and by integrating those amplitude-scaled typical spectra one obtains the threshold photon peak fluxes of 4.03\,ph\,cm$^{-2}$s$^{-1}$ for sGRB (or fluence of $3.92\times10^{-7}$\,erg\,cm$^{-2}$ for 256\,ms) and 2.22\,ph\,cm$^{-2}$s$^{-1}$ for lGRB (or fluence of $3.48\times10^{-7}$\,erg\,cm$^{-2}$ for 1024\,ms). These thresholds together with the expected duty cycle can be used to estimate an approximate number of GRB detections per year from the distribution of GRB photon peak fluxes from the FERMIGBRST catalog.

\textit{Fermi}/GBM surveys the entire sky, that is not occulted by the Earth, with the observing duty cycle of $\sim 85$\,\% \cite{Goldstein2019}. Following the trapped particle maps the duty cycle would be 76\,\% for orbital inclination of $89^\circ$ (polar orbit is an option for \textit{CAMELOT} CubeSats) and integral particle flux $\leq10$\,particle\,cm$^{-2}$s$^{-1}$ (see Sec.~\ref{sec:trapped} and Ref.~\citenum{Ripa2020}). However, we examined background data measured by a GRB instrument \textit{Lomonosov}/BDRG at polar LEO and it suggests that due to high background variation the duty cycle can be expected to be lower, i.e. about $60$\,\%. Therefore we assume this more conservative value of $\sim 60$\,\% as a duty cycle for a single \textit{CAMELOT} CubeSat. From these calculations we obtain an approximate number of sGRBs detectable by a single \textit{CAMELOT} CubeSat, i.e. with photon peak flux higher than the aforementioned thresholds, to be 18/year. In case of lGRBs we obtained 115/year.

In the same way we proceeded with SGRs. By using the simulated detection count rate of a typical SGR for the aluminium detector casing thickness of 1\,mm from Table~\ref{table:count_rates_transients} and the typical SGR spectral parameters calculated in Sec.~\ref{sec:SGR} we obtained SGR detection threshold photon flux of 9.19\,ph\,cm$^{-2}$s$^{-1}$ or threshold fluence of $8.47\times10^{-8}$\,erg\,cm$^{-2}$ for 0.2\,s and energy range of $15-500$\,keV. This means that all bursts listed in the Konus catalog of SGRs between 1978 and 2000 \cite{Aptekar2001} would be detectable also by a \textit{CAMELOT} CubeSat.

A difficulty is to estimate the SGR annual detection rate using this catalog because it was composed of measurements from several interplanetary spacecrafts and one LEO satellite which means it is difficult to know the exact duty cycle for the SGR measurements. Therefore we examined the five year \textit{Fermi}/GBM SGR catalog \cite{Collazzi2015} and calculated the detection thresholds also for the time scale of 0.1\,ms and energy range of $8-200$\,keV which are the median SGR duration and energy range used in this catalog. For these conditions we obtained \textit{CAMELOT} SGR photon flux detection threshold of 18.2\,ph\,cm$^{-2}$s$^{-1}$ or fluence threshold of $5.82\times10^{-8}$\,erg\,cm$^{-2}$. Using this fluence threshold, the SGR fluence distribution reported in the \textit{Fermi}/GBM SGR catalog and the aforementioned assumed duty cycle of a single \textit{CAMELOT} CubeSat we obtained a prediction of 46 SGRs detectable by one \textit{CAMELOT} satellite annually. Note that the annual number of detected SGRs will be subject to large fluctuations since SGR bursts tend to occur in clusters when particular magnetars become active.

Concerning the used SNR calculation it should be noted that it is rather simplistic for TGF detection. The large number of 0.1\,ms intervals (large number of trials) during the mission examined by the trigger algorithm (note that the rate trigger algorithm for \textit{CAMELOT} is yet under development) as well as the fact that a cosmic ray could easily cause a count in two detectors needs to be taken into account. If other background sources by chance produce one or two additional counts in the same 0.1\,ms interval, then a false trigger would be issued.

Specific conditions of the trigger algorithm to efficiently detect TGFs by \textit{CAMELOT} CubeSats are yet to be determined. For example one option is to aim to detect brighter but less frequent TGFs. \textit{AGILE}/MCAL observes 2780 TGFs within 3.5 years which is $\sim 800$ TGFs/year \cite{Maiorana2020}. The fluence distribution follows a power low of $-$2.2 \cite{Tierney2013}. This means that TGFs with 5 times greater fluence will be 3\,\% in number. The spectrum of a typical TGF used in our simulations is based on the TGF fluence at the threshold level of \textit{AGILE}/MCAL. Therefore if a \textit{CAMELOT} satellite had equtorial orbit as \textit{AGILE} then this scaling would give 23 TGFs/year/CubeSat with 19.5\,cnts/TGF.

However, \textit{CAMELOT} satellites will likely have polar or other high-inclination orbits. From observations most, if not all, TGFs has been detected at latitude lower than $45^\circ$ north and south by \textit{CGRO}/BATSE, \textit{RHESSI}, \textit{AGILE}/MCAL, \textit{Fermi}/GBM and ASIM instruments \cite{Ostgaard2019, Ursi2017}. One of the good distribution maps \cite{Ostgaard2019} was obtained from the recent ASIM instrument \cite{Neubert2019}. As relatively young (operational for $\sim 2$ years yet), the TGF number is not high, but with 51.6$^\circ$ on \textit{ISS}, it has a relatively uniform coverage of the TGF positional distribution. There are three major TGF sites: around Central America, around Central Africa and around South East Asia. A good comparison of the expected TGF detections by \textit{CAMELOT} can be done with the \textit{TARANIS}/XGRE instrument which was supposed to operate on Sun-synchronous orbit with 700\,km altitude. According to Ref.~\citenum{Sarria2017} \textit{TARANIS} (unfortunately lost due to the VEGA launch failure) was expected to detect $\sim 200$ TGFs/year. A \textit{CAMELOT} satellite will have about 1/10 of the effective area, but if corrected for the 500\,km vs. 700\,km altitude difference, this would be converted into $\sim 1/5$. Then $200 \times 3$\,\% (for $5\times$ brighter TGFs) give $\sim 6$ TGFs/year for a single satellite. With 9 satellites in a constellation CAMELOT would provide $\sim 50$ TGFs/year.

As shown in Fig.~\ref{fig:diff_direction_major_axis}, the \textit{CAMELOT} detectors can observe gamma-ray sources also for directions when the photons need to pass through the body of the satellite (rear direction). This means that also gamma-ray photons scattered off the Earth's atmosphere and entering the detector from the side not facing the source can produce signal. In this sense the \textit{CAMELOT} satellites will have omnidirectional FOV although the sensitivity for the rear direction is lower. Compton scatter of the burst flux off the Earth's atmosphere into the detector is known effect and observed already by the \textit{CGRO}/BATSE instrument \cite{Pendleton1996}. Correction for this effect has been included in the BATSE's response matrices \cite{Paciesas1999} and in the trigger efficiency calculation \cite{Pendleton1998}. See also Refs.~\citenum{Wigger2008, Wigger2006} with spectrum of GRB 021206 measured by the RHESSI satellite which shows significant Earth's atmospheric backscatter of photons below 300\,keV. Moreover, a method which employs the atmospheric scattering of GRB flux for the polarisation measurements in the prompt gamma-ray emission has been published in Ref.~\citenum{Willis2005} and Ref.~\citenum{McConnell1996}. The atmospheric scattering might affect the \textit{CAMELOT} measurements and it might be necessary to do careful modeling of this effect in the future in order to reduce the systematic uncertainties. A detailed simulation of this effect is beyond the scope of this paper, however such an analysis might be useful to improve the timing based localization in LEO.

\section{Conclusions}
\label{sec:conclusions}
 
A Geant4 based simulation was developed to understand the capabilities of the planned \textit{CAMELOT} CubeSat constellation to detect short and long GRBs, TGFs. The CAD model of the satellite was imported directly to Geant4 with CADMesh. Since the scintillators onboard \textit{CAMELOT} have a considerably large size, optical light propagation is important. To take this into account, scintillation light propagation was simulated in Geant4 by tracking each optical photon created by scintillation.
 
The simulation was validated and its optical parameters were calibrated with an $^{241}$Am X-ray source. The calibrated reflectivity of the surface of the scintillator turned out to be 99.99\,\% and absorption length 60\,cm.
 
Thirteen background components were simulated to determine their contribution to the overall background spectrum. The five components which contribute the most to background are: CXB (1000\,cps), albedo X-rays (200\,cps), cosmic-ray $\alpha$ particles (49\,cps), albedo protons (44\,cps) and albedo positrons (27\,cps). These count rates were calculated by assuming a 20\,keV low-energy threshold and 1\,mm of aluminium detector support structure thickness.
 
The total simulated background rate was 1545\,cps for a detector casing thickness of 0.5\,mm. By increasing the casing thickness to 1\,mm the total background decreased to 1410\,cps and by increasing it more to 1.5\,mm it turned out to be 1270\,cps. Finally for 2\,mm it was 1100\,cps.
These rates were obtained from models of fluxes of gamma-rays and particles averaged over various latitudes, depending on the particular flux component, below $50^\circ$ and outside SAA, because our goal was to obtain an expected ``mean" background rate in parts of orbit which are suitable for gamma-ray transient scientific data collection.
 
Since the four scintillators of the \textit{CAMELOT} CubeSat are placed on its two sides the direction of the source influences the signal-to-noise ratio. A typical short GRB was simulated in 70 directions. 35 directions were investigated by rotating the satellite around its major axis by 10$^{\circ}$ between simulations. The SNR of the detection of the typical short GRB (with integral fluxes between 8.15\,ph\,cm$^{-2}$s$^{-1}$ and 2.21\,ph\,cm$^{-2}$s$^{-1}$ for $E>5$\,keV and for 64 and 1024 ms integration window respectively) varied between 5 and $\sim10$. Other 35 directions were simulated by rotating the satellite around its minor axis. This resulted in less favorable directions, since the cross section with respect to the direction of X-rays from the sGRB is much smaller in this case. An SNR of at least 5 was determined for the range from 50$^{\circ}$ to 150$^{\circ}$.

The simulations show that CubeSats equipped with large area scintillators are able to detect sGRBs, lGRBs, TGFs and SGRs. In our case for the \textit{CAMELOT} CubeSats an average sGRB (256\,ms peak spectrum) could be detected with an SNR of $>12$ in the most favoured direction. lGRBs (with an integral flux of 2.54\,ph\,cm$^{-2}$s$^{-1}$ for $E>5$\,keV) yield an SNR of $>16$ for 4096\,ms exposure. TGFs despite their very short duration of 0.1\,ms could also be detected because, for example, for 1\,mm thicknesses of the Al detector casing the expected detected number of counts within 0.1\,ms from a TGF (with fluence at the threshold level of \textit{AGILE}/MCAL) is 3.9\,cnt, whereas the expected number of background counts is only 0.14\,cnt. SGRs due to their very large X-ray flux yield in an SNR of $>$100. The results of the simulation will aid the development of the trigger algorithm and also choosing the detector support structure.

% \disclosures 
%\subsection*{Disclosures}

\acknowledgments 

The research has been supported by the European Union, co-financed by the European Social Fund (Research and development activities at the E\"{o}tv\"{o}s Lor\'{a}nd University's Campus in Szombathely, EFOP-3.6.1-16-2016-00023). This work was partially supported by the GINOP-2.3.2-15-2016-00033 project which is funded by the Hungarian National Research, Development and Innovation Fund together with the European Union. This research was partially supported by JSPS and HAS under Japan - Hungary Research Cooperative Program. The research has been also supported by the Lend\"{u}let LP2016-11 grant awarded by the Hungarian Academy of Sciences. The authors would like express their sincere gratitude to Martino Marisaldi for the fruitful discussions on TGFs. The useful remarks of the anonymous referees are kindly acknowledged. Also the authors would like to thank at last but not least the engineers at C3S LLC for the support they provided with the CAD model of the satellite.

\newpage

\section{Appendix}
\label{appendix_a}

\begin{table}[h!]
\caption{The mass ratio of materials that are used for the satellite (Courtesy of C3S LLC).}
\begin{center}
\begin{tabular}{ |c|c|c|c|} 
 \hline
 Name of module & mass [g] & Type of material & Mass ratio [\%]\\\hline
 ADCS &	710	& Aluminum 6061-T6 &	50\\
			& & Copper Electric	& 25 \\
			& &  Glass Borosilicate 	& 25\\\hline
COM	& 	90	& 	Stainless Steel 	& 2\\
			& 	& Brass Generic		& 25\\
			& 	& Aluminum 7075-T73		& 40\\
			& 	& FR4 Glass-Epoxy sheet	& 	33\\\hline
EPS	& 	750	& 	FR4 Glass-Epoxy sheet		& 25\\
			& 	& Aluminum 6061-T6		& 75\\\hline
OBC	&	50		& FR4 Glass-Epoxy sheet	& 	100\\\hline
STRU	& 	980		& Aluminum 6061-T6	& 	100\\\hline
SP	& 	570		& Solar Panel	& 	100\\\hline
Payload	& 	100	& 	Aluminum 7075-T73	& 	100\\
 \hline
\end{tabular}
\end{center}
\label{table:materials1}
\end{table}

\begin{table}[h!]
\caption{The chemical composition of materials in mass fraction that are used for the satellite (Courtesy of C3S LLC).}
\begin{center}
\begin{tabular}{ |c|c|c|c|c|c|c|c|c|c|c|c|c|} 
 \hline
Material name & &&&&&&&&&&& \\\hline
Aluminum 6061-T6 &	Al & 96.90 &	Mg &	1.20 &	Si &	0.80 &	Fe &	0.70 &	Cu &	0.40 & &\\\hline		
Aluminum 7075-T73 &	Al &	88.60 &	Zn &	6.10 &	Mg &	2.90 &	Cu &	2.00 &	Si &	0.40 & &\\\hline		
Stainless Steel &	Fe &	66.50 &	Cr &	20.00 &	Ni &	10.50	&Mn &	2.00 &	Si &	1.00 & &\\\hline		
Copper Electric  &Cu &	100.00 & & & & & & & & & &	\\\hline			
Glass Borosilicate &	Si &	42.10 &	O &	54.80 &	B &	3.10 & & & & & &\\\hline			
FR4 Glass-Epoxy &	Si &	23.39 &	O &	36.02 &	C &	37.04 &	H &	3.55 & & & &\\\hline		
Brass Generic &	Cu &	85.00 &	Zn &	15.00 & & & & & & & &\\\hline						
Solar Panel &	Ge &	38.00 &	Si &	24.00 &	O &	20.00 &	C &	13.00 &	H &	4.00 &	B &	1.00\\\hline
\end{tabular}
\end{center}
\label{table:materials2}
\end{table}

\begin{figure}[h]
\begin{center}
\begin{tabular}{c}
\includegraphics[width=0.8\linewidth]{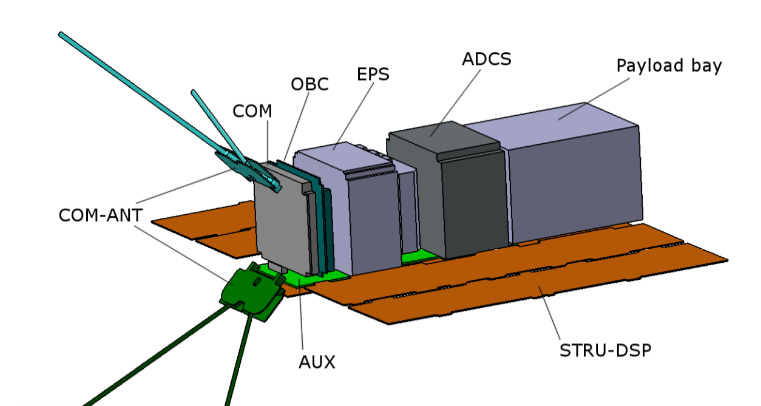}
\end{tabular}
\end{center}
\caption 
{ \label{fig:themodel}
The individual volumes of the simulated satellite. The CAD model of the satellite was read into Geant4 and 4 scintillators (not displayed on this figure) with their respective read out were placed on two sides of the satellite (Courtesy of C3S LLC).} 
\end{figure}

\newpage

%%%%% References %%%%%

\bibliography{main}   % bibliography data in main.bib
\bibliographystyle{spiejour}   % makes bibtex use spiejour.bst

%%%% Biographies of authors %%%%%

%\listoffigures
%\listoftables

\end{spacing}
\end{document}